\documentclass[12pt]{article}
\usepackage{latexsym}
\usepackage{bm}
\usepackage{amsbsy}
\usepackage[authoryear]{natbib}

\def\R{{\rm I \mkern-2.5mu \nonscript\mkern-.5mu R}}

\usepackage{float}
\usepackage{graphicx}

\newcommand{\be}{\begin{equation}}
\newcommand{\ee}{\end{equation}}
\newcommand{\bea}{\begin{eqnarray}}
\newcommand{\eea}{\end{eqnarray}}
\newcommand{\bean}{\begin{eqnarray*}}
\newcommand{\eean}{\end{eqnarray*}}

\renewcommand{\theequation}{\arabic{section}.\arabic{equation}}

\newcommand{\sect}[1]{\section{#1} \setcounter{equation}{0}
                      \setcounter{table}{0} \setcounter{figure}{0}}

\newcommand{\Exp}{{\rm I\hspace{-0.7mm}E}}

\newcommand{\vect}[1]{\pmb{#1}}
\newcommand{\mat}[1]{\boldsymbol{\bm #1}}

\title{Likelihood estimation for stochastic differential
  equations with mixed effects}

\author{Fernando Baltazar-Larios \\
{\small Facultad de Ciencias \vspace{-1.5mm}} \\
{\small Universidad Nacional Aut\'onoma de M\'exico 
\vspace{-1.5mm}} \\ 
{\small A.P.\ 20-726, 01000 Mexico, D.F., Mexico\vspace{-1.5mm}} \\
{\small fernandobaltazar@ciencias.unam.mx}
\and 
Mogens Bladt \\
{\small Deptartment of Mathematical Sciences \vspace{-1.5mm}} \\
{\small University of Copenhagen\vspace{-1.5mm}} \\
{\small Universitetsparken 5, DK-2100 Copenhagen {\O}, 
Denmark\vspace{-1.5mm}} \\ 
{\small bladt@math.ku.dk}
\and
Michael S\o rensen \\ 
{\small Deptartment of Mathematical Sciences \vspace{-1.5mm}} \\
{\small University of Copenhagen\vspace{-1.5mm}} \\
{\small Universitetsparken 5, DK-2100 Copenhagen {\O}, 
Denmark\vspace{-1.5mm}} \\ 
{\small michael@math.ku.dk}}

\date{}

\begin{document}

\maketitle

\begin{abstract}
Stochastic differential equations provide a powerful
tool for modelling dynamic phenomena affected by random noise. In case
of repeated observations of time series for several experimental
units, it is often the case that some of the parameters vary between
the individual experimental units, which has motivated a considerable
interest in stochastic differential equations with mixed effects,
where a subset of the parameters are random. These models enable
simultaneous representation of randomness in the dynamics and
variability between experimental units. When the data are observations
at discrete time points, the likelihood function is only rarely
explicitly available, so for likelihood-based inference to be feasible,
numerical methods are needed. We present Gibbs samplers and stochastic
EM-algorithms based on augmented data obtained by the simple method
for simulation of diffusion bridges in Bladt and Sørensen (2014). This
method is easy to implement and has no tuning parameters. The method
is, moreover, computationally efficient at low sampling frequencies
because the computing time increases linearly with the time between
observations. The algorithms can be extended to models with
measurement errors. The Gibbs sampler as well as the EM-algorithm are
shown to simplify considerably for exponential families of diffusion
processes, including many models used in practice. In a simulation
study, the estimation methods are shown to work well for
Ornstein-Uhlenbeck processes and t-diffusions with mixed
effects. Finally, the Gibbs sampler is applied to neuronal data. 
\end{abstract}

\sect{Introduction}

Stochastic differential equations provide a useful and versatile tool for
modelling dynamic phenomena affected by random noise, which has been
applied in several sciences. The data are time series of observations
at discrete time points. In case of repeated observations of time
series for several experimental units, for instance observations for
different individuals or at 
different locations, it is often the case that some of the parameters
vary between the individual experimental units. This is particularly the case for
biological data, but is also common in the social sciences. Various
types of mixed models have for a long time been used widely in biostatistics and
in econometrics, where they are called models for panel data.

As a statistical tool for such repeated measurements of dynamical data, 
there has been considerable interest in stochastic differential equations with
mixed effects, i.e.\ where a subset of the parameters are
random, with independent values drawn for each experimental unit,
while the rest are population parameters that are common to all
observations. This approach has several advantages. The random effects
model intersubject/unit variations, which may be of interest in
themselves, but also implies that several time series can be analysed 
simultaneously, because the random effects take care of model
uncertainty and environmental variation. This can substantially
increase the statistical power, and therefore, for instance, provides
a way around the well known problem that estimators of parameters in the
drift coefficient are often highly imprecise because the time series
is too short. These parameters can be well determined by observing a
sufficient number of repeated time series, which is often much easier to
obtain than longer time series. Stochastic differential equations
with mixed effects have been applied to several areas of the
biological sciences. Applications to neuro science,
electroencephalography, pharmacokinetics and growth of animals and
tumors can, for instance, be found in \cite{umberto1}, \cite{wiqvist},
\cite{ruse}, \cite{samson2}, \cite{samson3}, \cite{umberto4} and \cite{braumann2}.

The price for the advantages of using stochastic differential equations
with mixed effects is the computational effort needed to
obtain maximum likelihood or Bayesian  estimators. While the dynamics
is formulated in continuous time, the observations are made at
discrete points in time, which implies that there is rarely an explicit
expression for the likelihood function, even in the case of fixed
effects. A further complication for random effects models is that the
fixed effects likelihood 
function must be integrated with respect to the distribution of the
random effects. Even when the fixed effects likelihood function
is explicit this can rarely be done analytically. Therefore numerical methods or
analytical approximations must be used to obtain estimators. 

Several approaches have been studied. For high frequency data
relatively simple approximations are available. One approach is to use
the fixed effects likelihood function (conditional on the random
effects) obtained from the Girsanov formula in the
hypothetical situation, where the diffusion sample paths have been
observed continuously in an interval. Estimators and theory are then
developed for continuous time data, while in practice the estimators
are approximated by replacing the integrals in the Girsanov formula
by Riemann and Ito sums. Some of the papers give bounds on
the error made by these approximations in terms of the sampling
frequency. This approach was taken by \cite{samson1,samson4,samson5} 
and \cite{ruse}, who considered exponential families of diffusion
processes in the sense of \cite{kuso}, where the distribution of the
random effects is the conjugate prior or a finite mixture of such
distributions. In this case the likelihood function for continuous
time data can be calculated explicitly. A disadvantage of the approach
based on the continuous time likelihood function is that it only works
when there are neither fixed parameters nor random effects in the
diffusion coefficient. This limitation is avoided in another method
for high frequency data proposed by \cite{laredo1,laredo2}, where
the conditional fixed effects likelihood function is approximated by
the pseudo-likelihood obtained from the Euler approximation to the
distribution of the discrete time observations. Like in the previous
papers, these authors considered exponential families of diffusion
processes with conjugate priors as the distribution of the random
effects in order to obtain an explicit 
pseudo-likelihood. Nonparametric estimation for high frequency data
was investigated by \cite{comte}, \cite{dion2} and \cite{dion1}, while
high frequency asymptotics was studied by \cite{maitra} and
\cite{delattre2}  and in papers referenced in these papers. A different
approach to asymptotics based on local asymptotic normality and
$L^2$-differentiability was taken in \cite{ruse}.

When the sampling frequency is not sufficiently high, better
approximations to the likelihood function are needed. \cite{umberto2}
and \cite{umberto3} used the Hermite polynomial expansions of the fixed effect
likelihood function by \cite{ait-sahalia} combined with Gaussian
quadrature or Laplace's method to approximate the
likelihood function numerically. This can be rather time-consuming, so a
number of other methods have been proposed. \cite{madsen} obtained an
approximation to the likelihood function via the extended Kalman
filter, while \cite{braumann1} used the Delta method (i.e.\ a second
order Taylor expansion) to obtain an approximation to the fixed effect
likelihood function that can be integrated analytically with respect to
many distributions of the random effects. 

More precise approximations can be obtained by thinking of the data as
a missing data problem, where the 
missing data are the random effects for the individual experimental
units, the sample paths between the observation times and possibly
measurement errors. For the full augmented data set the likelihood
function is explicitly available. This suggests application of the EM
algorithm or the Gibbs sampler to obtain the maximum likelihood estimator
or to simulate from the posterior. \cite{samson2, samson6}
used the stochastic approximation EM algorithm (SAEM) combined with
the Metropolis-Hastings algorithm or particle filters with sample paths
simulated by the Euler scheme on a fine grid. \cite{delattre1}
moreover used the extended Kalman filter to reduce the amount of
simulation needed at the cost of loosing control of the error
resulting from the linearization needed for non-linear diffusion
models. The Gibbs sampler was used by \cite{samson3},
\cite{golightly}, \cite{umberto4} and \cite{wiqvist} in combination
with a generalisation of the approximate diffusion bridge proposed by
\cite{durhamgallant}, the pseudo-marginal Metropolis-Hastings
algorithm and particle filters or synthetic likelihoods.

In recent years random-effects models based on stochastic
differential equations driven by fractional Brownian motion and
partial stochastic differential equations have been studied, see
\cite{elomari1}, \cite{dai}, \cite{bishwal}, \cite{prakasarao} and
\cite{elomari2}.  

In the present paper, we also augment of the data and apply both a
stochastic EM algorithm and the Gibbs sampler. We propose to combine
these algorithms with the simple method for simulation of diffusion bridges
introduced by \cite{bladtsorensen}; see also the corrigendum 
\cite{bladtmidersorensen}. Advantages over other methods is that it is
easy to implement and that there are no tuning parameter that must be
set appropriately (except the discretization of the Euler scheme, but
this is the case for all methods and is not difficult to tune). Another
advantage is that our method is efficient for low frequency observations, as
the computing time increase only linearly with the time between observations.
To simplify the presentation we consider
only one-dimensional diffusions proceses, but our algorithms can be
directly generalised to multivariate diffusions the applying
the methods for simulating multivariate diffusion bridges in   
\cite{bladtfinchsorensen,corbladtfinchsorensen}.

The paper is organized as follows. The model and the data augmentation
are presented in Section \ref{model}, and the Gibbs sampler and the
stochastic EM algorithm are introduced for general models in Section
\ref{MCMCEM}. In Section \ref{errors} it is outlined how the
algorithms can be straightforwardly extended to include the case of
measurement errors, and in Section \ref{exponential} it is investigated how the
algorithms simplify when the fixed effects model obtained conditionally on
the random effects is an exponential family of diffusion processes,
which is the case for many models used in practice. It is investigated how
well the new methods work for the Ornstein-Uhlenbeck process and for
a $t$-diffusion in simulation studies in Section \ref{simul}, and in
Section \ref{neurodata} the Gibbs sampler is applied to neuronal
data. The simple method for diffusion bridge simulation is briefly
explained in Appendix \ref{appen}.

\sect{Model and augmented data}
\label{model}

Consider $N$ diffusion processes
\be 
\label{basicmodel}
dX^i_t = d_{\vect{\alpha}, \vect{a}^i}(X^i_t)dt+\sigma_{\vect{\beta},
  \vect{b}^i} (X^i_t)dW^i_t , \ \ \ i=1,\ldots,N
\ee
where $W^i$, $i=1,\cdots, N$ are independent standard Wiener
processes. The vectors $\vect{\alpha}$ and $\vect{\beta}$ are parameters to be
estimated, while the vectors $\vect{a}^i$ and $\vect{b}^i$ are random effects. The
random vectors $(\vect{a}^i, \vect{b}^i)$, $i=1,\ldots,N$, are independent,
identically distributed random vectors with density
$p_{\vect{\gamma}}(\vect{a}, \vect{b})$ with respect to some dominating
measure, and  they are independent of the Wiener processes $W^{i}$,
$i=1,\ldots,N$. Thus the parameters to be estimated are $\vect{\theta} = 
(\vect{\alpha}, \vect{\beta}, \vect{\gamma})$. We assume that
$\sigma_{\vect{\beta}, \vect{b}} (x)>0$  for all $x$ in the state
interval and for all values of $(\vect{\beta}, \vect{b})$, that
$d_{\vect{\alpha}, \vect{a}}(x)$ is continuously differentiable
w.r.t. $x$ for all values of $(\vect{\alpha}, \vect{a})$, and that
$\sigma_{\vect{\beta}, \vect{b}}(x)$  is twice continuously
differentiable w.r.t.\ $x$ for all $(\vect{\beta}, \vect{b})$. Moreover, 
we assume that for any given value of $(\vect{a}^i, \vect{b}^i)$ the
stochastic differential equation (\ref{basicmodel}) has a unique weak
solution, and that the speed measure is finite.

We consider discrete time data $\vect{X}_{\mbox{\tiny obs}}  =
(\vect{X}^1_{\mbox{\tiny obs}},\ldots, \vect{X}^N_{\mbox{\tiny obs}})$, where
\[
\vect{X}^i_{\mbox{\tiny obs}} = \left( x^i_1, \ldots x^i_{n_i}\right),
\]
and $x^i_j = X^i_{t^i_j}$, with $t^i_1<t^i_2<...<t^i_{n_i}$.

There is usually no explicit expression for the likelihood function for
the discrete time data $\vect{X}_{\mbox{\tiny obs}}$ conditional on the
random effects $(\vect{a}^i,\vect{b}^i)$, $i=1,\ldots,N$, but the data
$\vect{X}_{\mbox{\tiny obs}}$ are partial observations of an augmented data
set consisting of the random effects $(\vect{a}^i,\vect{b}^i)$, $i=1,\ldots,N$ and
the complete continuous time observations of the  diffusion processes
$X^1, \ldots, X^N$, i.e.\ observations of $X^i$ in the time interval
$[t^i_1, t^i_{n_i}]$, $i=1,\ldots,N$. To augment the data, we must simulate the
missing data. In particular, we need to simulate each of the processes
$X^{i}$ conditionally on $\vect{X}^i_{\mbox{\tiny obs}}$ and on the
random effects. This we can do by simulating independent $(t^i_{j-1},
x^i_{j-1},t^i_j, x^i_j)$-bridges, $j=1,\ldots,n_i$, for the diffusion
(\ref{basicmodel}) with fixed values of $\vect{a}^i$ and $\vect{b}^i$. The process
obtained by conditioning on $X_{t_2}=x_2$ for a solution of
(\ref{basicmodel}) in the interval $[t_1,t_2]$ with $X_{t_1}=x_1$ is
called a $(t_1, x_1, t_2, x_2)$-bridge. We propose to simulate the
diffusion bridges by the simple method introduced in
\cite{bladtsorensen} and the corrigendum \cite{bladtmidersorensen}.
The algorithm is briefly described in Appendix \ref{appen}. Advantages
of this method are is that it is easy to understand and implement, and
that it is efficient at low sampling frequencies, because the
computing time increases linearly with $t_2-t_1$.

If the diffusion coefficient
depends on $\vect{\beta}$ and $\vect{b}$, then the probability
measures corresponding to continuous time observations of the diffusion
model given by (\ref{basicmodel}) are singular (for different values of
the $\vect{\beta}$ and $\vect{b}$), so that the likelihood function for this
augmented data set does not exist. Therefore, we use the Lamperti transformation 
\be
\label{h-trans}
h_{\vect{\beta},\vect{b}}(x)=\int_{x^{\ast}}^x
\frac{1}{\sigma_{\vect{\beta},\vect{b}} (y)}dy, 
\ee
where $x^{\ast}$ is a point in the state-interval. In the following we
assume that $\vect{b}^i$ is fixed (i.e.\ we argue conditionally on the random
vector $\vect{b}^i$). By Ito's formula, the process
  $Y^i_t=h_{\vect{\beta},\vect{b}^i}(X^i_t)$ solves  
\be
\label{Y} 
dY^i_t=\mu_{\vect{\alpha}, \vect{\beta}, \vect{a}^i, \vect{b}^i}(Y^i_t)dt+dW^i_t,  
\ee
where the drift coefficient is
\be
\label{Y-drift}
\mu_{\vect{\alpha}, \vect{\beta}, \vect{a},
  \vect{b}}(y)=\frac{d_{\vect{\alpha},\vect{a}}(h_{\vect{\beta},\vect{b}}^{-1}(y))} 
{\sigma_{\vect{\beta},\vect{b}}(h_{\vect{\beta},\vect{b}}^{-1}(y))}-\frac{1}{2} 
\sigma_{\vect{\beta},\vect{b}}^{\prime}\left(h_{\vect{\beta},\vect{b}}^{-1}(y)\right),
\ee
with $\sigma_{\vect{\beta},\vect{b}}'(x) = \partial_x \sigma_{\vect{\beta},\vect{b}}(x)$.
In (\ref{Y}) the diffusion coefficient does not depend on the
parameters and random effects, which is a necessary condition for the
existence of the likelihood function. The basic dominating measure for
the likelihood function is the  Wiener measure induced on
$C([t_1,t_{n_i}])$ by the standard Wiener 
process. It is an assumption in the rest of the paper that the
measures induced on $C([t_1,t_{n_i}])$ by $Y^i$ (for all values of the
parameters and the random effects) are dominated by the Wiener
measure, and that the Radon-Nikodym derivatives are given by
Girsanov's theorem.  

The Lamperti transformation (\ref{h-trans}), and hence the transformed
data, depends on the random effect $\vect{b}^i$ and the parameter $\vect{\beta}$.
To reduce the Fisher information in the missing data and thus make the
MCMC and EM algorithms more efficient, we follow
\cite{roberts} and augment the discrete time data in the following way
that involves an extra path transformation, see also
\cite{beskosetal}. Our augmented  ``full''  data set is $(\vect{X}_{\mbox{\tiny obs}},
\vect{X}_{\mbox{\tiny mis})}$, where $\vect{X}_{\mbox{\tiny mis}} =
(\vect{X}^1_{\mbox{\tiny mis}}, \ldots, \vect{X}^N_{\mbox{\tiny mis}})$ and 
\[
\vect{X}^i_{\mbox{\tiny mis}} = \{ \vect{Y}^{*i}, \vect{a}^i, \vect{b}^i \}.
\]
Here $\vect{Y}^{*i}$ = $\{ Y^{*ij}_t, \, t \in [t^i_{j-1},t^i_j], \,
j=2,\ldots,n_i \}$ with
\[
Y^{*ij}_t = Z^{ij}_t -
\ell^{ij}_{\vect{\beta},\vect{b}^i}(t), \ \ \ t \in [t^i_{j-1},t^i_j],
\] 
where
\be
\label{g-trend}
\ell^{ij}_{\vect{\beta},\vect{b}^i}(t) =
\frac{(t^i_j-t)h_{\vect{\beta},\vect{b}^i}(x^i_{j-1})
+ (t-t^i_{j-1})h_{\vect{\beta},\vect{b}^i} (x^i_j)}
{t^i_j-t^i_{j-1}}, \hspace{5mm} t \in  [t^i_{j-1}, t^i_j],
\ee
$j=2, \ldots, n_i$, interpolates linearly between the Lamperti
transformed discrete time observations, and where, conditionally on
$\vect{X}^i_{\mbox{\tiny obs}}$ and  
$(\vect{a}^{i}, \vect{b}^i)$, the processes $Z^{ij}_t$, $j=2,
\ldots, n_i, i=1, \ldots, N$ are independent 
$(t^i_{j-1}, h_{\vect{\beta},\vect{b}^{i}}(x^i_{j-1}),t^i_j,$ 
$h_{\vect{\beta},\vect{b}^{i}}(x^i_j))$-bridges for the diffusion $Y^i$ given by
(\ref{Y}) with parameter values $(\vect{\alpha},\vect{\beta})$, and $(\vect{a}^{i},
\vect{b}^i)$. The bridge $Z^{ij}$ can be obtain from a $(t^i_{j-1}, x^i_{j-1},t^i_j,  
x^i_j)$-bridge, $X^{ij}$, of the basic diffusion given by (\ref{basicmodel})
by the transformation $Z^{ij}_t = h_{\vect{\beta},\vect{b}^{i}}(X^{ij}_t)$. 
The process $Y^{*ij}$ is not in general a $(t^i_{j-1},0, t^i_j,
0)$-bridge. Under the dominating Wiener measure, however, $Y^i$ is a
Brownian motion, so $Y^{*ij}$ is a Brownian $(t^i_{j-1},0, t^i_j, 0)$-bridge. This
is needed in the derivation of the following expression for the likelihood
function.

The likelihood function for the augmented data set is
\[
L(\vect{\alpha}, \vect{\beta}, \vect{\gamma} ; \vect{X}_{\mbox{\tiny obs}},
\vect{X}_{\mbox{\tiny mis}}) = \prod_{i=1}^N L_i(\vect{\alpha}, \vect{\beta};
\vect{X}^i_{\mbox{\tiny obs}}, \vect{X}^i_{\mbox{\tiny mis}})  
\prod_{i=1}^N p_{\vect{\gamma}}(\vect{a}^i,\vect{b}^i),
\]
where
\bean
\lefteqn{\log L_i(\vect{\alpha}, \vect{\beta}; \vect{X}^i_{\mbox{\tiny
      obs}}, \vect{X}^i_{\mbox{\tiny mis}}) = 
  H_{\vect{\alpha}, \vect{\beta}, \vect{a}^i,
    \vect{b}^i}(x^i_1, x^i_{n_i})}   
\\ && \\
&& \hspace{5mm}
- \sum_{j=2}^{n_i} \left[
\frac{(h_{\vect{\beta},\vect{b}^i}(x^i_j)-h_{\vect{\beta},
    \vect{b}^i}(x^i_{j-1}))^2}{2(t^i_j-t^i_{j-1})}    
+ \log (\sigma_{\vect{\beta},\vect{b}^i} (x^i_j)) 
+\frac{1}{2} \int_{t^i_{j-1}}^{t^i_j}  
\phi_{\vect{\alpha}, \vect{\beta}, \vect{a}^i, \vect{b}^i}(Y^{*ij}_s +
\ell^{ij}_{\vect{\beta},\vect{b}^i}(s)) ds 
\right].  
\eean
Here $h_{\vect{\beta}, \vect{b}}$ is given by (\ref{h-trans}),
\bean
\phi_{\vect{\alpha}, \vect{\beta}, \vect{a}, \vect{b}}(x) &=&
\mu_{\vect{\alpha}, \vect{\beta},\vect{a},\vect{b}}^{\prime}(x) 
+\mu_{\vect{\alpha}, \vect{\beta},\vect{a},\vect{b}}(x)^2 \\
&=& d'_{\vect{\alpha},\vect{a}}(h_{\vect{\beta},\vect{b}}^{-1}(x))-
2d_{\vect{\alpha},\vect{a}}(h_{\vect{\beta},\vect{b}}^{-1}(x))
\frac{\sigma'_{\vect{\beta},\vect{b}}(h_{\vect{\beta},\vect{b}}^{-1}(x))}
{\sigma_{\vect{\beta},\vect{b}}(h_{\vect{\beta},\vect{b}}^{-1}(x))} \\
&& \ \ - \mbox{\small $\frac12$}
\sigma''_{\vect{\beta},\vect{b}}(h_{\vect{\beta},\vect{b}}^{-1}(x))
\sigma_{\vect{\beta},\vect{b}}(h_{\vect{\beta},\vect{b}}^{-1}(x))
+\mbox{\small $\frac14$}
\left(\sigma'_{\vect{\beta},\vect{b}}(h_{\vect{\beta},\vect{b}}^{-1}(x))
\right)^2 +
\frac{d^2_{\vect{\alpha},\vect{a}}(h_{\vect{\beta},\vect{b}}^{-1}(x))}
{\sigma^2_{\vect{\beta},\vect{b}}(h_{\vect{\beta},\vect{b}}^{-1}(x))},
\eean
and
\be
\label{g} 
H_{\vect{\alpha}, \vect{\beta}, \vect{a}, \vect{b}}(x,y) = 
\int^{h_{\vect{\beta},\vect{b}}(y)}_{h_{\vect{\beta},\vect{b}}(x)}
  \mu_{\vect{\alpha}, \vect{\beta}, \vect{a}, \vect{b}}(y) dy  =
\int_{x}^y \frac{d_{\vect{\alpha},\vect{a}}(y)}{\sigma_{\vect{\beta}, 
\vect{b}}^2(y)}dy -\mbox{\small $\frac12$} 
\log \left(
  \frac{\sigma_{\vect{\beta},\vect{b}}(y)}{\sigma_{\vect{\beta},
\vect{b}}(x)} \right).
\ee
The expression for $L_i(\vect{\alpha}, \vect{\beta}; \vect{X}^i_{\mbox{\tiny obs}},
\vect{X}^i_{\mbox{\tiny mis}})$ follows from Girsanov's theorem and Ito's
formula by arguments in \cite{roberts}.

\sect{MCMC and EM algorithms}
\label{MCMCEM}

In order to use the likelihood function for the augmented data to find
the maximum likelihood estimator, or an analogous Bayesian estimator,
for the discrete time data, we can apply the EM-algoritm or the Gibbs
sampler. In this section we present a Gibbs sampler and a stochastic
EM-algorithm. To clarify the general structure of the algorithms, they
are presented for the general model defined above. In section
\ref{exponential} we consider a particularly tractable sub-class of
models for which the algorithms simplify considerably.

\subsection{The Gibbs sampler}

For MCMC-estimation, we first specify a suitable prior
$\pi(\vect{\alpha},\vect{\beta},\vect{\gamma})$. Then we can simulate
from the posterior distribution of $\vect{\theta} = (\vect{\alpha}, \vect{\beta},
\vect{\gamma})$ by means of the following Gibbs sampler with $2N+1$
sites. The sites are $\vect{\theta}$
plus $(\vect{a}^i,\vect{b}^i)$ and $\vect{Y}^{*i}$ for
$i=1,\ldots,N$. As we shall see later, it can be preferable to have
more sites, for instance, $\vect{\alpha}$, $\vect{\beta}$,
$\vect{\gamma}$, $\vect{a}^i$ and $\vect{b}^i$, $i=1,\ldots,N$. As a
bi-product we also obtain a sample of values of the random effects
conditional on the observations.

To start the sampler, draw $\vect{\theta}$ from the prior $\pi$, and given the value of
$\vect{\gamma}$, draw $(\vect{a}^i, \vect{b}^i)$ from the distribution with  
density function $p_{\vect{\gamma}}$, independently for
$i=1,\ldots,N$. Finally, simulate independent sample paths $Y^{*ij}$ conditionally on
  $\vect{\alpha}, \vect{\beta}, \vect{a}^i, \vect{b}^i$ and
  $\vect{X}^i_{\mbox{\tiny obs}}$ \vspace{2mm} for $j=2, \ldots, n_i,
  \ i=1, \ldots, N$. Then repeat the following algorithm.

\begin{description}

\item{1.} Draw $\vect{\theta}$ conditionally on $(\vect{X}_{\mbox{\tiny mis}},
  \vect{X}_{\mbox{\tiny obs}})$

\item{2.} For $i=1,\ldots,N$, draw independent values of
  $(\vect{a}^i,\vect{b}^i)$ conditionally on $(\vect{\theta}, \vect{X}^i_{\mbox{\tiny obs}}, \vect{Y}^{*i})$    

\item{3.} Simulate independent sample paths $Y^{*ij}$ conditionally on
 $(\vect{\alpha}, \vect{\beta}, \vect{a}^i, \vect{b}^i,
 \vect{X}^i_{\mbox{\tiny obs}})$ \vspace{2mm} for $j=2, \ldots, n_i,
 \ i=1, \ldots, N$ 

\item{4.} GO TO 1

\end{description}

In step 1 and 2 the conditional densities of $\vect{\theta}$ and $(\vect{a}^i,
\vect{b}^i)$ are proportional to
$\pi(\vect{\theta}) \cdot L(\vect{\theta} ;
\vect{X}_{\mbox{\tiny  obs}}, \vect{X}_{\mbox{\tiny mis}})$ and
$L_i(\vect{\alpha}, \vect{\beta}; \vect{X}^i_{\mbox{\tiny obs}},
\vect{Y}^{*i},\vect{a}^i,\vect{b}^i)
p_{\vect{\gamma}}(\vect{a}^i,\vect{b}^i)$,
respectively. In Section \ref{exponential}  
we shall se that these steps simplify for an important class of
diffusion models, but for more complicated models, it is usually
necessary to use Metropolis within Gibbs. 

To simulate $Y^{*ij}$ in step 3, we must simulate a $(t^i_{j-1},
h_{\vect{\beta},\vect{b}_{i}}(x^i_{j-1}),t^i_j, h_{\vect{\beta},\vect{b}_{i}}(x^i_j))$-bridge for
the diffusion $Y^i$ given by (\ref{Y}) (or equivalently a $(t^i_{j-1}, x^i_{j-1},t^i_j,  
x^i_j)$-bridge of the diffusion given by (\ref{basicmodel})). We do
this using the method by \cite{bladtsorensen}, see 
also \cite{bladtmidersorensen}, which is briefly presented in the
appendix. The main reasons for this choice are 
easy implementation and the fact that the computing time for this
method is linear in the length of the interval $[t^i_{j-1}, t^i_j]$
and hence works for all sampling frequencies. The simplest method is
to use the approximate diffusion bridges obtain by a simple rejection
sampler, which usually has a high acceptance rate. In our experience
this approximation is sufficiently accurate to obtain good
estimates. Exact diffusion bridges can be obtained by Metropolis
within Gibbs, as explained in the appendix. In the $m$th iteration we
simulate, as a proposal, an approximate $(t^i_{j-1}, x^i_{j-1},t^i_j,  
x^i_j)$-bridge, $X^{m,ij}$, of the diffusion (\ref{basicmodel})) supplemented
by its associated geometric variable $S_{m,ij}$. The proposed bridge is accepted
with probability $\min \{1, S_{m,ij}/S_{m-1,ij} \}$. Otherwise we keep
the bridge used in iteration $m-1$.

\subsection{The EM algorithm}

An alternative to MCMC-estimation is the EM-algorithm, which under
weak conditions converges to a (possibly local) maximum of the
likelihood function for the discrete time data, see e.g.\ \cite{mclachlan}. \\

\noindent
The EM-algorithm works as follows. 
\begin{description}
\item{0.} Choose initial values $\hat{\vect{\alpha}}_0,
  \hat{\vect{\beta}}_0, \hat{\vect{\gamma}}_0 $, $k:=0$    
\item{1.} (E--step) Calculate the function
\[
Q(\vect{\alpha}, \vect{\beta}, \vect{\gamma}) = \sum_{i=1}^N
\Exp_{\hat{\vect{\alpha}}_k, \hat{\vect{\beta}}_k,
  \hat{\vect{\gamma}}_k}  \left[  \left. \log L_i(\vect{\alpha},
  \vect{\beta};  \vect{X}^i_{\mbox{\tiny obs}},
  \vect{X}^i_{\mbox{\tiny mis}}) + \log p_{\vect{\gamma}}(\vect{a}^i,
  \vect{b}^i) \, \right| \, \vect{X}^i_{\mbox{\tiny obs}} \right].     
\]
\item{2.} (M--step) $(\hat{\vect{\alpha}}_{k+1},
  \hat{\vect{\beta}}_{k+1}, \hat{\vect{\gamma}}_{k+1}) 
=\mbox{argmax}_{\vect{\alpha}, \vect{\beta}, \vect{\gamma}}
Q(\vect{\alpha}, \vect{\beta}, \vect{\gamma})$. 
\item{3.} k:=k+1; GO TO 1. 
\end{description}

\vspace{3mm}

In the E--step it is almost always impossible to calculate the conditional
expectations of the log-likelihood functions explicitly. Therefore we
calculate the $i$th conditional expectation numerically by generating MC-samples 
$\vect{X}^{m,i}_{\mbox{\tiny mis}} = \{ Y^{*m,i},  \vect{a}^{m,i},
\vect{b}^{m,i} \} $, $m=1,\ldots,M$ (conditionally
on the observed data $\vect{X}^i_{\mbox{\tiny obs}}$) 
such that we can approximate $Q(\vect{\alpha}, \vect{\beta}, \vect{\gamma})$ by
$\hat Q(\vect{\alpha}, \vect{\beta}, \vect{\gamma}) = \sum_{i=1}^N \hat
Q_i(\vect{\alpha}, \vect{\beta}, \vect{\gamma}) $, where 
\[
\hat Q_i(\vect{\alpha}, \vect{\beta}, \vect{\gamma}) := \frac1M \sum_{m=1}^M \left\{ 
\log L_i(\vect{\alpha}, \vect{\beta} ; \vect{X}^i_{\mbox{\tiny obs}},
\vect{Y}^{*m,i}, \vect{a}^{m,i}, \vect{b}^{m,i}) +\log
p_{\vect{\gamma}}(\vect{a}^{m,i}, \vect{b}^{m,i})  \right\}. 
\]
This can be done by means of the following simplified version of the previous
Gibbs sampler.

\begin{description}

\item{0.} Draw $(\vect{a}^i,\vect{b}^i)$ from \vspace{2mm} $p_{\hat{\vect{\gamma}}_k}$

\item{1.} Simulate independent sample paths $Y^{*ij}$, $j=2, \ldots,
  n_i$, conditionally on \vspace{2mm} $\vect{a}^i, \vect{b}^i,
  \vect{X}^i_{\mbox{\tiny obs}}$ (with the parameter values
  $\hat{\vect{\alpha}}_k$ and $\hat{\vect{\beta}}_k$) 

\item{2.} Draw $(\vect{a}^i, \vect{b}^i)$ conditionally on $\vect{X}^i_{\mbox{\tiny obs}},
  \vect{Y}^{*i}$ (with the parameter values
  $\hat{\vect{\alpha}}_k$, $\hat{\vect{\beta}}_k$ and $\hat{\vect{\gamma}}_k$) 

\item{3.} GO TO 1

\end{description}

In step 1 of the Gibbs sampler, we use again the simple method by
\cite{bladtsorensen}, see also \cite{bladtmidersorensen}, to simulate
diffusion bridges, cf.\ the appendix. 
Again there is a choice between using approximate bridges or using Metropolis
within Gibbs to obtain exact diffusion bridges, as explained previously.
In step 2, the conditional density of $(\vect{a}^i, \vect{b}^i)$ is proportional
to $L_i(\hat{\vect{\alpha}}_k, \hat{\vect{\beta}}_k ;
\vect{X}^i_{\mbox{\tiny obs}}, \vect{Y}^{*i},\vect{a}^i,\vect{b}^i)  
p_{\hat{\vect{\gamma}}_k}(\vect{a}^i, \vect{b}^i)$. As for the
previous Gibbs sampler, it is often necessary to use Metropolis within
Gibbs for complicated models.

When the conditional expectation of the log-likelihood function for
the full data set is calculated by Monte Carlo methods, as we do, the algorithm
is often referred to Monte Carlo EM (MCEM), see \cite{weitanner}.
In our experience this works well without computational problems. 
If desired, the MCEM algorithm can be replaced by a stochastic EM
algorithm, see e.g.\ \cite{dieboltip}, or a stochastic approximation
EM (SAEM), see \cite{delyonlaviellemoulines}, both of which are easily
implemented using the simple method for diffusion bridge
simulation.  

\sect{Measurement errors}
\label{errors}

In several applications, e.g.\ pharmacokinetics, measurement errors
are expected. In this section we will outline how it is straightforward
to extend our methods to include measurement errors. 

Assume that the observations are $\vect{U}_{\mbox{\tiny obs}}  =
(\vect{U}^1_{\mbox{\tiny obs}},\ldots, \vect{U}^N_{\mbox{\tiny obs}})$, where
$\vect{U}^i_{\mbox{\tiny obs}} = \left( u^i_1, \ldots u^i_{n_i}\right)$,
and $$u^i_j = X^i_{t^i_j} + \varepsilon^i_j,$$ with
$t^i_1<t^i_2<...<t^i_{n_i}$. Here the process $X^i_t$ is defined as in
Section \ref{model}, and the random variables $\varepsilon^i_j \sim
N(0,\tau^2)$ are mutually independent and independent of $X^i,
\vect{a}^i, \vect{b}^i$, $i = 1, \ldots , N$. We consider the
case of additive Gaussian measurement errors, but the general case
where $u^i_j = k(X^i_{t^i_j}, \varepsilon^i_j)$ for some function $k$
with the variables  $\varepsilon^i_j $ non-Gaussian, can be treated in
exactly the same way -- apart from a more complicated notation.

Here the unobserved data include the measurement errors, i.e.\ the
augmented data set is $(\vect{U}_{\mbox{\tiny obs}},  
\vect{U}_{\mbox{\tiny mis})}$ with $\vect{U}_{\mbox{\tiny mis}} =
(\vect{U}^1_{\mbox{\tiny mis}}, \ldots, \vect{U}^N_{\mbox{\tiny mis}})$ and 
$\vect{X}^i_{\mbox{\tiny mis}} = \{ \vect{Y}^{*i}, \vect{a}^i,
\vect{b}^i, \vect{\varepsilon}^i \}$. Here $\vect{\varepsilon}^i =
(\varepsilon^i_1, \ldots , \varepsilon^i_{n_i} )$, $Y^{*ij}_t = \tilde{Z}^{ij}_t -
\tilde{\ell}^{ij}_{\vect{\beta},\vect{b}^i, \vect{\varepsilon}^i}(t),
\ t \in [t^i_{j-1},t^i_j]$ and 
\[
\tilde{\ell}^{ij}_{\vect{\beta},\vect{b}^i, \vect{\varepsilon}^i }(t) =
\frac{(t^i_j-t)h_{\vect{\beta},\vect{b}^i}(u^i_{j-1}-\varepsilon^i_{j-1})
+ (t-t^i_{j-1})h_{\vect{\beta},\vect{b}^i} (u^i_j-\varepsilon^i_j)}
{t^i_j-t^i_{j-1}}, \hspace{3mm} t \in  [t^i_{j-1}, t^i_j],
\]
$j=2, \ldots, n_i$. Moreover, $\tilde{Z}^{ij}_t =
h_{\vect{\beta},\vect{b}^{i}}(X^{ij}_t)$, where $X^{ij}$, $j=2, 
\ldots, n_i, i=1, \ldots, N$ are independent $(t^i_{j-1},
u^i_{j-1}-\varepsilon^i_{j-1}, t^i_j, u^i_j-\varepsilon^i_j )$-bridges
of the basic diffusion given by (\ref{basicmodel}) conditionally on 
$\vect{U}^i_{\mbox{\tiny obs}}$, $(\vect{a}^{i}, \vect{b}^i)$ and
$\vect{\varepsilon}^i$. The likelihood function for the augmented data set is
\[
L(\vect{\alpha}, \vect{\beta}, \vect{\gamma}, \tau^2; \vect{U}_{\mbox{\tiny obs}},
\vect{U}_{\mbox{\tiny mis}}) = \prod_{i=1}^N \left[ \tilde{L}_i(\vect{\alpha}, \vect{\beta};
\vect{U}^i_{\mbox{\tiny obs}} - \vect{\varepsilon}^i, \vect{X}^i_{\mbox{\tiny mis}})  
p_{\vect{\gamma}}(\vect{a}^i,\vect{b}^i) \prod_{j=1}^{n_i}
\varphi_{\tau^2}(\varepsilon^i_j) \right], 
\]
where $\varphi_{\tau^2}$ is the Gaussian density function with mean
zero and variance $\tau^2$ and $\tilde{L}_i$ is defined as $L_i$ in
Section \ref{model} except that $\ell^{ij}_{\vect{\beta},\vect{b}^i}$ is replaced by
$\tilde{\ell}^{ij}_{\vect{\beta},\vect{b}^i, \vect{\varepsilon}^i}$.

Apart from the modified likelihood function and obvious minor changes,
the only two changes to the Gibbs sampler are that in step 1 also
$\tau^2$ must be drawn, and that in each iteration an
extra step must be added prior to step 3. In the extra step,
$\vect{\varepsilon}^i$ is simulated 
conditionally on $(\vect{\alpha}, \vect{\beta}, \vect{\gamma}, \tau^2,
\vect{X}^i_{\mbox{\tiny  obs}}, \vect{Y}^{*i}, \vect{a}^i,\vect{b}^i)$ 
for $i=1,\ldots,N$. For the EM-algorithm similar modifications are
needed. In particular, $\vect{\varepsilon}^i$ must be added as an
extra site to the Gibbs sampler used to create the MC-samples.   

The algorithms could also have been formulated in terms of simulation
of generalized diffusion bridges, i.e.\ diffusion bridges where the
two endpoints are random, in this case $u^i_{j-1}-\varepsilon^i_{j-1}$
and $u^i_j-\varepsilon^i_j $.

\sect{Exponential family models}
\label{exponential}

Considerable simplifications of the Gibb's sampler and the
EM-algorithm can be obtained, when the drift is linear in the
parameter $\vect{\alpha} \in \R^{p_1}$  and in the random effect
$\vect{a} \in \R^{p_2}$, and the random effects $\vect{a}$ and
$\vect{b}$ are independent with densities $p_{\vect{\gamma}_1}$ and
$p_{\vect{\gamma}_2}$, respectively. Specifically, we will in this
section investigate stochastic differential equations with mixed
effects, where the drift has the form
\be
\label{lindrift}
 d_{\vect{\alpha}, \vect{a}}(x) = \sum_{k=1}^{p_1} \alpha_k f_k(x) +
 \sum_{k=1}^{p_2} a_k g_k(x) = \vect{\alpha} \vect{f}(x)^\top + \vect{a}
 \vect{g}(x)^\top , 
\ee
where $\vect{f} = (f_1, \ldots, f_{p_1})$ and $\vect{g} = (g_1,
\ldots, g_{p_2})$. In this paper vectors are row vectors, and $\top$
denotes transposition. A large part of the diffusion models used in
practice have this form.

When the drift is of the form (\ref{lindrift}), the continuous time full model is an
exponential family of processes (in the sense of \cite{kuso}) in the
parameters and random effects in the drift (conditionally on these
random effects). More specifically, we find that
\be
\label{expH}
H_{\vect{\alpha},\vect{\beta}, \vect{a}, \vect{b}}(x,y) = \vect{\alpha}
\int_{x}^y \frac{\vect{f}(y)^\top}{\sigma^2_{\vect{\beta},\vect{b}}(y)}dy  
+ \vect{a} \int_{x}^y \frac{\vect{g}(y)^\top}{\sigma^2_{\vect{\beta},\vect{b}}(y)}dy 
-\mbox{\small $\frac12$} \log \frac{\sigma_{\vect{\beta},
\vect{b}}(y)}{\sigma_{\vect{\beta},\vect{b}}(x)},
\ee
cf.\ (\ref{g}), and
\bean
\phi_{\vect{\alpha},\vect{\beta}, \vect{a}, \vect{b}}(x) &=&
\vect{\alpha} \bar{\vect{f}}_{\vect{\beta},\vect{b}}(h_{\vect{\beta},
  \vect{b}}^{-1}(x))^\top + \vect{a}
\bar{\vect{g}}_{\vect{\beta},\vect{b}}(h_{\vect{\beta},\vect{b}}^{-1}(x))^\top 
+ 2 \vect{\alpha}
\tilde{\vect{f}}_{\vect{\beta},\vect{b}}(h_{\vect{\beta},\vect{b}}^{-1}(x))^\top 
\tilde{\vect{g}}_{\vect{\beta},\vect{b}}(h_{\vect{\beta},\vect{b}}^{-1}(x))
\vect{a}^\top \\ && \\ 
&& + \vect{\alpha}
\tilde{\vect{f}}_{\vect{\beta},\vect{b}}(h_{\vect{\beta},\vect{b}}^{-1}(x))^\top 
\tilde{\vect{f}}_{\vect{\beta},\vect{b}}(h_{\vect{\beta},\vect{b}}^{-1}(x))
\vect{\alpha}^\top 
+ \vect{a} \tilde{\vect{g}}_{\vect{\beta},\vect{b}}(h_{\vect{\beta},\vect{b}}^{-1}(x))^\top
\tilde{\vect{g}}_{\vect{\beta},\vect{b}}(h_{\vect{\beta},\vect{b}}^{-1}(x))
\vect{a}^\top \\ && \\ 
&& + \mbox{\small $\frac14$} \left(
  \sigma'_{\vect{\beta},\vect{b}}(h_{\vect{\beta},\vect{b}}^{-1}(x))\right)^2 
-\mbox{\small $\frac12$}
\sigma''_{\vect{\beta},\vect{b}}(h_{\vect{\beta},\vect{b}}^{-1}(x))
\sigma_{\vect{\beta},\vect{b}}(h_{\vect{\beta},\vect{b}}^{-1}(x)), 
\eean
where $h_{\vect{\beta}, \vect{b}}$ is given by (\ref{h-trans}), and
the $k$th coordinate of the functions $\bar{\vect{f}}, 
 \bar{\vect{g}},  \tilde{\vect{f}}$ and $\tilde{\vect{g}}$ are given by
 \bean
(\bar{\vect{f}}_{\vect{\beta},\vect{b}})_k(x)  &=&  f'_k(x) - 2f_k(x)(\log
(\sigma_{\vect{\beta},\vect{b}}(x))' \\
(\bar{\vect{g}}_{\vect{\beta},\vect{b}})_k(x)  &=& g'_k(x) - 2g_k(x)(\log
(\sigma_{\vect{\beta},\vect{b}}(x))' \\
(\tilde{\vect{f}}_{\vect{\beta},\vect{b}})_k(x) &=&
f_k(x)/\sigma_{\vect{\beta},\vect{b}}(x) \\ 
(\tilde{\vect{g}}_{\vect{\beta},\vect{b}})_k(x) &=&
g_k(x)/\sigma_{\vect{\beta},\vect{b}}(x). 
\eean

If we assume that the distribution of the random effects in the
drift is Gaussian
\be
\label{distra}
  \vect{a}^i \sim N_{p_2}(\vect{\xi}, \mat{\Gamma}^{-1}),
\ee
then the likelihood function for the augmented data has the form
\bea
\label{exponentiallikelihood1}
\hspace{-10mm}
L(\vect{\theta}; \vect{X}_{\mbox{\tiny obs}},
\vect{X}_{\mbox{\tiny mis}}) &=& \hspace{-1mm}  \exp \left( \vect{\alpha}
  \vect{v}^\top_{\vect{\beta}, \underline{\vect{a}}, \underline{\vect{b}}}
  - \mbox{\small $\frac12$}
  \vect{\alpha} \mat{D}_{\vect{\beta}, \underline{\vect{b}}} \, \vect{\alpha}^\top +
  q(\vect{\beta}, \underline{\vect{a}}, \underline{\vect{b}}) \right) 
\prod_{i=1}^N \left[p_{\vect{\gamma}_1}(\vect{a}^i)
  p_{\vect{\gamma}_2}(\vect{b}^i) \right] \\
&=& \hspace{-1mm}  C(\vect{\theta}, \underline{\vect{b}})
\prod_{i=1}^N \exp \left( \vect{a}^i \left[ (\vect{t}_{\vect{\alpha},
    \vect{\beta}, \vect{b}^i}^i)^\top 
 \hspace{-1mm} + \mat{\Gamma} \vect{\xi}^\top \right] - \mbox{\small $\frac12$}
  \vect{a}^i\left( \mat{B}_{\vect{\beta}, \vect{b}^i}^i + \mat{\Gamma}
  \right) \! (\vect{a}^i)^\top \right) \hspace{-1mm}. 
\label{exponentiallikelihood2}
\eea
Here we have used the notation $\underline{\vect{a}} = (\vect{a}^1, \ldots,
\vect{a}^N)$ and  $\underline{\vect{b}} = (\vect{b}^1, \ldots,\vect{b}^N)$.
The likelihood function is Gaussian as a function of $\vect{\alpha}$
as well as of $\vect{a}^i$. The vectors and matrices in
(\ref{exponentiallikelihood1}) and (\ref{exponentiallikelihood2}) are
defined as follows: 
\bea
\label{matA}
\vect{t}_{\vect{\alpha}, \vect{\beta}, \vect{b}^i}^i&=&\int^{x^i_{n_i}}_{x^i_1}
\frac{\vect{g}(y)}{\sigma^2_{\vect{\beta},\vect{b}^i}(y)}dy 
  -\sum_{j=2}^{n_i}  \int_{t_{j-1}^i}^{t_j^i} \vect{k}_{1i}(Y^{*ij}_s +
  \ell_{\vect{\beta},\vect{b}^i}^{ij}(s))ds \\
  \label{matK}
\vect{v}_{\vect{\beta}, \underline{\vect{a}}, \underline{\vect{b}}}
&=&\sum_{i=1}^N \left\{ \int^{x^i_{n_i}}_{x^i_1} 
  \frac{\vect{f}(y)}{\sigma^2_{\vect{\beta},\vect{b}^i}(y)}dy 
  -\sum_{j=2}^{n_i}  \int_{t_{j-1}^i}^{t_j^i} \vect{k}_{2i}(Y^{*ij}_s +
  \ell_{\vect{\beta},\vect{b}^i}^{ij}(s))ds \right\} \\
  \label{matB}
  \mat{B}_{\vect{\beta}, \vect{b}^i}^i&=&\sum_{j=2}^{n_i}  \int_{t_{j-1}^i}^{t_j^i}
  \vect{K}_{3i}(Y^{*ij}_s +\ell_{\vect{\beta},\vect{b}^i}^{ij}(s)) ds \\
\label{matD}
\mat{D}_{\vect{\beta}, \underline{\vect{b}}}&=&\sum_{i=1}^N \sum_{j=2}^{n_i}
\int_{t_{j-1}^i}^{t_j^i} 
\vect{K}_{4i}(Y^{*ij}_s + \ell_{\vect{\beta},\vect{b}^i}^{ij}(s))ds
\eea
where
\bean
\vect{k}_{1i}(y) &=&  \mbox{\small $\frac12$}
  \bar{\vect{g}}_{\vect{\beta},\vect{b}^i}(h_{\vect{\beta},\vect{b}^i}^{-1}(y))
  +\vect{\alpha}
  \tilde{\vect{f}}_{\vect{\beta},\vect{b}^i}(h_{\vect{\beta},\vect{b}^i}^{-1}(y))  ^\top  
  \tilde{\vect{g}}_{\vect{\beta},\vect{b}^i}(h_{\vect{\beta},\vect{b}^i}^{-1}(y)) \\
\vect{k}_{2i}(y) &=& \mbox{\small $\frac12$} 
  \bar{\vect{f}}_{\vect{\beta},\vect{b}^i}(h_{\vect{\beta},\vect{b}^i}^{-1}(y)) 
+ \vect{a}^i
\tilde{\vect{g}}_{\vect{\beta},\vect{b}^i}(h_{\vect{\beta},\vect{b}^i}^{-1}(y)) ^\top 
\tilde{\vect{f}}_{\vect{\beta},\vect{b}^i}(h_{\vect{\beta},\vect{b}^i}^{-1}(y)) \\
\vect{K}_{3i}(y) &=&
\tilde{\vect{g}}_{\vect{\beta},\vect{b}^i}(h_{\vect{\beta},\vect{b}^i}^{-1}(y))^\top 
\tilde{\vect{g}}_{\vect{\beta},\vect{b}^i}(h_{\vect{\beta},\vect{b}^i}^{-1}(y))  \\ 
\vect{K}_{4i}(y) &=&
\tilde{\vect{f}}_{\vect{\beta},\vect{b}^i}(h_{\vect{\beta},\vect{b}^i}^{-1}(y))^\top 
\tilde{\vect{f}}_{\vect{\beta},\vect{b}^i}(h_{\vect{\beta},\vect{b}^i}^{-1}(y))
\eean
The real functions $q(\vect{\beta}, \underline{\vect{a}}, \underline{\vect{b}})$ and
$C(\vect{\theta}, \underline{\vect{b}})$
can be determined from the expressions above (and
the multivariate normal density function). 

\subsection{The Gibbs sampler}
\label{expGibbs}

For the Gibbs sampler the simplification is obtained if a conjugate prior is used
for $\vect{\alpha}$ and for $\vect{\gamma}_1$. Specifically, we use the priors
\bean
\vect{\alpha} &\sim& N_{p_1}(\bar{\vect{\alpha}}, \mat{\Sigma}) \\
\vect{\gamma}_1 & = & (\vect{\xi},\mat{\Gamma}) \sim
NW_{p_2}(\vect{\xi}_0, \lambda, \mat{V}, \nu) \\
\vect{\beta} &\sim& \pi_1 \hspace{5mm} \vect{\gamma}_2 \sim \pi_2 
\eean
 where $\vect{\alpha}$, $\vect{\beta}$, $\vect{\gamma}_1$ and $\vect{\gamma}_2$
are independent. By $NW_{p_2}(\vect{\xi}_0, \lambda, \mat{V}, \nu)$ we
denote the $p_2$-dimensional normal-Wishart distribution with
parameters $\vect{\xi}_0 \in \R^{p_2}, \lambda > 0 , \mat{V}, \nu >
p_2-1$, where $\mat{V}$ is a positive definite $p_2 \times p_2-$matrix.
The parameter $\vect{\gamma}_1 = (\vect{\xi},\mat{\Gamma})$ can be
simulated by first 
simulating $\mat{\Gamma}$ from a Wishard distribution with parameters
$(\mat{V}, \nu)$, and then (conditionally on $\mat{\Gamma}$)
simulating $\vect{\xi}$ from a multivariate normal distribution with
mean $\vect{\xi}_0$ and covariance matrix $(\lambda \mat{\Gamma})^{-1}$.

If some of the parameters or random effects must 
necessarily be positive (for instance to ensure ergodicity), then
these coordinates of the multivariate normal distributions must be
restricted to the positive half-line. The same is then the case of the
corresponding normal distributions in the following algorithm.

With these priors, it follows easily from (\ref{exponentiallikelihood1})
and (\ref{exponentiallikelihood2}) (and well-known results for the
normal-Wishart prior) that the Gibbs sampler \vspace{2mm}  goes as follows.

\begin{description}

\item{0.}  To start the sampler, draw $\vect{\theta} = (\vect{\alpha}, \vect{\beta},
\vect{\gamma}_1, \vect{\gamma}_2)$ from the prior distribution,
and given the value of $(\vect{\gamma}_1, \vect{\gamma}_2)$, draw
$(\vect{a}^i,\vect{b}^i)$ from the distribution with parameter
$(\vect{\gamma}_1, \vect{\gamma}_2)$, independently for
$i=1,\ldots,N$. Complete the initialization by simulating independent
sample paths $Y^{*ij}$ conditionally on $\vect{\theta}, \vect{a}^i,
\vect{b}^i$ and $\vect{X}^i_{\mbox{\tiny obs}}$ for
$j=2, \ldots, n_i, \ i=1, \ldots, N$. 

\item{1.} Draw $\vect{\alpha}$ from the
  $N_{p_1}((\vect{v}_{\vect{\beta}, \underline{\vect{a}}, \underline{\vect{b}}} +\bar{\vect{\alpha}} \mat{\Sigma}^{-1})
  (\mat{D}_{\vect{\beta}, \underline{\vect{b}}}
  +\mat{\Sigma}^{-1})^{-1}, (\mat{D}_{\vect{\beta},
    \underline{\vect{b}}} +\mat{\Sigma}^{-1})^{-1})$ 
  distribution (conditionally on
  $\vect{\beta}, \underline{\vect{a}}, \underline{\vect{b}}, \vect{X}_{\mbox{\tiny
      obs}}$ and $\vect{Y}^{*i}$, $i=1, \ldots, N$) 
 
\item{2.} Draw $\vect{\beta}$ from the distribution with density
  function proportional to $$\pi_1(\vect{\beta})\prod_{i=1}^N
  L_i(\vect{\alpha}, \vect{\beta}; \vect{X}^i_{\mbox{\tiny obs}},
  \vect{Y}^{*i}, \vect{a}^i,\vect{b}^i)$$ (all
  quantities other than $\vect{\beta}$ are fixed)  

\item{3.} Draw $\vect{\gamma}_1 = (\vect{\xi}, \vect{\Gamma})$ 
(conditionally on $\vect{a}^1,\ldots,\vect{a}^N$)  from the \\
$NW_{p_2}(\lambda \vect{\xi}_0 + N \bar{\vect{a}})/(\lambda+N), \lambda + N,  
(\mat{V}^{-1}+ N \hat{\mat{\Gamma}}^{-1} + (\bar{\vect{a}} -
\vect{\xi}_0) (\bar{\vect{a}} - \vect{\xi}_0)^\top\lambda 
N/(\lambda+N))^{-1}, \nu+N)$ \\ distribution, where $\bar{\vect{a}}$ and
$\hat{\vect{\Gamma}}^{-1}$ are the mean and the sample covariance
matrix of $\vect{a}^1,\ldots,\vect{a}^N$  

\item{4.} Draw $\vect{\gamma}_2$ from the distribution with density
  function proportional to $\pi_2(\vect{\gamma}_2)
  \prod_{i=1}^Np_{\vect{\gamma}_2}(\vect{b}^i)$ (with
  $\vect{b}^1,\ldots,\vect{b}^N$ fixed)

\item{5.} Draw independent values of $\vect{a}^i$ from the 
  $N_{p_2}((\vect{t}_{\vect{\alpha}, \vect{\beta},
    \vect{b}^i}^i+\vect{\xi}  \mat{\Gamma})(\mat{B}_{\vect{\beta}, \vect{b}^i}^i + 
  \mat{\Gamma})^{-1}, (\mat{B}_{\vect{\beta}, \vect{b}^i}^i
  +\mat{\Gamma})^{-1})$ distribution, 
  $i=1,\ldots,N$ (conditionally on $\vect{\theta},\vect{b}^i, \vect{X}^i_{\mbox{\tiny
      obs}}$ and $\vect{Y}^{*i}$)  

\item{6.} Draw $\vect{b}^i$ from the distribution with density function
  proportional to $$L_i(\vect{\alpha}, \vect{\beta};
  \vect{X}^i_{\mbox{\tiny obs}}, \vect{Y}^{*i}, \vect{a}^i,\vect{b}^i)
  p_{\vect{\gamma}_2}(\vect{b}^i)$$ (where all  
quantities other than $\vect{b}^i$ are fixed), independently for
$i=1,\ldots,N$

\item{7.} Simulate independent sample paths $Y^{*ij}$ conditionally on
  $\vect{\theta}, \vect{a}^i, \vect{b}^i$ and $\vect{X}^i_{\mbox{\tiny obs}}$
  \vspace{2mm}  for $j=2, \ldots, n_i, \ i=1, \ldots, N$ 

\item{8.} GO TO 1

\end{description}

In (\ref{distra}) we allowed all kinds of dependencies between the
random effects in the drift. It might in some cases be reasonable to assume some
structure in the matrix $\Gamma$, for instance that it is a diagonal
matrix. This changes the posterior of $\vect{\gamma}_1$ in step 3. If
we assume that the coordinates of $\vect{a}^i$ are independent, then the 
posterior is independent normal-gamma distributions. 

If some of the parameters or random effects must necessarily be
positive, the normal distributions, restricted to the positive
half-line, can be replaced by exponential distributions independent of
the other coordinates. For instance, if $a^i$ is one-dimensional
($p_2=1$) and must be positive, then we can assume that $a^i$ is
exponential distributed with mean 
$\gamma_1^{-1}$, and that the prior of $\gamma_1$ is the $\Gamma(\nu,
\lambda)$-distribution (with density proportional
to $\eta^{\nu-1} e^{-\lambda \eta}$). Then step 5 is replaced by
\begin{description}
\item{5*.}
For $i=1,\ldots,N$, draw independent values of $a^i$ 
from the normal distribution with mean $(t_{\vect{\alpha}, \vect{\beta}, \vect{b}^i}^i -\gamma_1)/B_{\vect{\beta}, \vect{b}^i}^i$ and variance
$(B_{\vect{\beta}, \vect{b}^i}^i)^{-1}$ restricted to the positive half-axis (conditionally on
$\vect{\theta}, \vect{b}^i,
\vect{X}^i_{\mbox{\tiny obs}}$ and $\vect{Y}^{*i}$), 
\end{description}
while step 3 is replaced by
\begin{description}
\item{3*.}
  Draw $\gamma_1$ from the $\Gamma (\nu +N, \lambda + a^1+\ldots +
  a^N)$-distribution (conditionally on $a^1,\ldots, a^N$).  
\end{description}

A considerable simplification of step 2 can be obtained when
\be
\label{sigmabeta}
\sigma_{\beta, \vect{b}}(x) = \beta c_{\vect{b}}(x),
\ee
where $\beta>0$ and $c_{\vect{b}}(x)>0$. Then
\[
H_{\vect{\alpha},\beta, \vect{a}, \vect{b}}(x,y) = \beta^{-2} \int_{x}^y
\frac{\vect{\alpha} \vect{f}(z)^\top +  \vect{a} \vect{g}(z)^\top}{c_{\vect{b}}^2(z)}dz
-\mbox{\small $\frac12$} \log \frac{c_{\vect{b}}(y)}{c_{\vect{b}}(x)},
\]
where the $\beta$ in the last term in (\ref{expH}) has been omitted as
it cancels in the likelihood function. Further, 
\[
  \phi_{\vect{\alpha},\beta,\vect{a}, \vect{b}}(x) = m^{(1)}_{\vect{\alpha},1, \vect{a}, \vect{b}}(\beta x)  
+ \beta^2 m^{(2)}_{1, \vect{b}}(\beta x) + \beta^{-2}
m^{(3)}_{\vect{\alpha},1, \vect{a}, \vect{b}}(\beta x) 
\]
where
\bea
\nonumber
m^{(1)}_{\vect{\alpha},\beta, \vect{a}, \vect{b}} (x) &=& \vect{\alpha}
\bar{\vect{f}}_{\beta,\vect{b}} (h_{\beta, \vect{b}}^{-1}(x))^\top +
\vect{a} \bar{\vect{g}}_{\beta,\vect{b}} (h_{\beta, \vect{b}}^{-1}(x))^\top \\
\label{mmm}
m^{(2)}_{\beta, \vect{b}} (x) &=& \mbox{\small $\frac14$} (\sigma_{\beta,\vect{b}}'(h_{\beta, \vect{b}}^{-1}(x)))^2
-\mbox{\small $\frac12$} \sigma_{\beta,\vect{b}}''(h_{\beta, \vect{b}}^{-1}(x)) 
\sigma_{\beta, \vect{b}}(h_{\beta, \vect{b}}^{-1}(x))
\\ \nonumber
m^{(3)}_{\vect{\alpha},\beta, \vect{a}, \vect{b}} (x) &=& 2
\vect{\alpha} \tilde{\vect{f}}_{\beta,\vect{b}} (h_{\beta,
  \vect{b}}^{-1}(x))^\top 
\tilde{\vect{g}}_{\beta,\vect{b}} (h_{\beta, \vect{b}}^{-1}(x)) \vect{a}^\top 
+ \vect{\alpha} \tilde{\vect{f}}_{\beta,\vect{b}} (h_{\beta, \vect{b}}^{-1}(x))^\top
\tilde{\vect{f}}_{\beta,\vect{b}} (h_{\beta, \vect{b}}^{-1}(x))
\vect{\alpha}^\top \\ \nonumber
&& \hspace{70mm} \mbox{} + \vect{a} \tilde{\vect{g}}_{\beta,\vect{b}} (h_{\beta, \vect{b}}^{-1}(x))^\top 
\tilde{\vect{g}}_{\beta,\vect{b}} (h_{\beta, \vect{b}}^{-1}(x)) \vect{a}^\top
\eea 

Thus, in step 2 of the Gibbs sampler, $\beta$ must be drawn from a
distribution with density proportional to
\[
\pi_1(\beta)\beta^{-(n_\cdot-N)}\exp\left( - \beta^{-2} (G_1 + G_2) + F(\beta)  \right),
\]
where $n_\cdot = n_1+ \cdots + n_N$ and
\[
G_1 = \sum_{i=1}^N \sum_{j=2}^{n_i}
\frac{(h_{1, \vect{b}^i} (x_j^i)-h_{1, \vect{b}^i} (x_{j-1}^i))^2}{2(t_j^i - t_{j-1}^i)} \hspace{8mm}
G_2 = - \sum_{i=1}^N \int_{x_1^i}^{x_{n_i}^i}
\frac{\vect{\alpha}\vect{f}(z)^\top + \vect{a}^i \vect{g}(z)^\top}{c_{\vect{b}^i}^2(z)}dz
\]
and 
\[
F(\beta) = 
- \frac12 \sum_{i=1}^N \sum_{j=2}^{n_i}
\int_{t_{j-1}^i}^{t_j^i} \phi_{\vect{\alpha},\beta, \vect{a^i}, \vect{b}^i}(Y^{*ij}_s +
  \beta^{-1}\ell_{1, \vect{b}^i}^{ij}(s)) ds,
\]
where the function $F$ in general depends on $\beta$ in a complicated
way.

If we introduce the parameter $\eta = \beta^{-2}$ and as the
prior of $\eta$ choose the $\Gamma( \kappa, \delta)$-distribution,
then in step 2, $\eta$ should be drawn from the weighted
gamma-distribution with density function proportional to 
\be
\label{weightedgamma}
\eta^{(n_\cdot-N)/2 + \kappa -1}\exp\left( - \eta (\delta + G_1 + G_2)
  + F(\eta^{-1/2})  \right).
\ee
This can be done in several ways. 

a) We can replace step 2 by one iteration of a Metropolis-Hastings
algorithm.

\begin{description}
\item{2*.} If $\delta + G_1 + G_2 > 0$, then draw $\eta^*$ from the
$\Gamma( (n_\cdot-N)/2 + \kappa, \delta + G_1 + G_2)$ distribution,
and with probability
\[
\min \left( 1, \exp(F(\eta^{*-1/2}) - F(\eta^{-1/2})) \right)
\]
accept the proposed value and set $\eta := \eta^*$.

If $\delta + G_1 + G_2  \leq0$, then draw $\eta^*$ from the
$\Gamma( (n_\cdot-N)/2 + \kappa, \delta + G_1)$ distribution,
and with probability
\[
\min \left( 1, \exp(F(\eta^{*-1/2}) - F(\eta^{-1/2}) + (\eta - \eta^*)G_2) \right)
\]
accept the proposed value  and set $\eta := \eta^*$.

If the proposed value is not accepted, $\eta$ is unchanged.
\end{description}

b) If the function $F$ is bounded, step 2 can be replaced by a rejection
sampler. Suppose, for instance, that the prior of $\eta$ is truncated
to the interval $[E_1, E_2]$, $0 < E_1 < E_2$, and that $F$ is a
continuous function of $\beta \in [E_2^{-1/2} ,E_1^{-1/2}]$. Then there exists
$M>0$ such that $\sup_{E_1 \leq \eta \leq E_2} \exp(F(\eta^{-1/2}))
\leq M$. If $\delta + G_1 + G_2 > 0$, 
we can replace step 2 by the following algorithm. 
\begin{description}
\item{2**.} 
\begin{description}
\item{(1)} Draw $\eta^*$ from the $\Gamma((n_\cdot-N)/2 + \kappa, \delta + G_1 +
G_2)$-distribution truncated to the interval $[E_1,E_2]$. 

\item{(2)} With probability $\exp(F(\eta^{*-1/2}))/M$, accept the
  proposed value and set $\eta := \eta^*$. Otherwise go to (1).
\end{description}
\end{description}
If $\delta + G_1 + G_2 \leq 0$, a similar step can be used, where the
scale parameter of the gamma distribution is modified as above, and the
acceptance probability is $\exp(F(\eta^{*-1/2}) - \eta^*G_2 )/M_2$
where $\exp(F(\eta^{-1/2})- \eta G_2)  \leq M_2$.

c) Finally, we can use {\it approximate direct sampling}. Again we consider the
case $\delta + G_1 + G_2 > 0$. For numerical reasons we choose $M$
such that $(G_1 + G_2)/M$ has a reasonable magnitude. Draw $Z_1,
\ldots, Z_K$ independently from the $\Gamma((n_\cdot-N)/2 + \kappa,
(\delta + G_1 +G_2)/M)$-distribution, and define  
\[
p_i := \frac{\exp(F((Z_i/M)^{-1/2}))}{\sum_{j=1}^K \exp(F((Z_j/M)^{-1/2}))}.
\]
Let $I$ be a random variable with $P(I = i) = p_i$. Then the
distribution of $\eta = Z_I/M$ is approximately equal to the weighted
gamma-distribution (\ref{weightedgamma}), where the approximation
improves as $K$ increases. Again there is a modification in case $\delta +
G_1 + G_2 \leq 0$ 

For concrete models, it may be more efficient to take advantage of the
particular structure of the function $F$, when drawing values of
$\beta$ in step 2 of the Gibbs sampler, see examples below.

Step 6 can be simplified in a similar way when
\be
\label{sigmab}
\sigma_ {\vect{\beta}, b}(x) = b c_{\vect{\beta}}(x),
\ee
where $b>0$ and $c_{\vect{\beta}}(x)>0$. For more details see the next
subsection.

\subsection{The EM algorithm}

For the EM algorithm we obtain a further simplification, if we assume
that the diffusion coefficient has the form (\ref{sigmab}) and that
the random variable $E = b^{-2}$ is $\Gamma (\kappa, \delta)$ distributed. Thus
$\vect{\gamma} = (\vect{\xi}, \mat{\Gamma}, \kappa, \delta)$, and the
function $q(\vect{\beta})$, implicitly given by
(\ref{exponentiallikelihood1}) (here we suppress the arguments
$\underline{\vect{a}}$ and $\underline{\vect{b}}$), has the form
\bea
q(\vect{\beta}) &=& \sum_{i=1}^N \left[ \frac{\vect{a}^i}{(b^i)^2} \int_{x^i_1}^{x^i_{n_i}}
  \frac{\vect{g}(z)^\top}{c^2_{\vect{\beta}}(z)} dz - (n_i-1) \log (b^i) \right. 
-\frac12 \log (c_{\vect{\beta}}(x^i_{n_i})/c_{\vect{\beta}}(x^i_1))
-\sum_{j=2}^{n_i} \log c_{\vect{\beta}}(x^i_j) \nonumber \\
&& \label{EMq} \mbox{} - \frac{1}{(b^i)^2} \sum_{j=2}^{n_i}
\frac{(h_{\vect{\beta},1}(x^i_{j}) - h_{\vect{\beta},1}(x^i_{j-1})
  )^2}{2(t_j^i - t^i_{j-1})} - \left. \frac12 \sum_{j=2}^{n_i} 
  \int_{t_{j-1}^i}^{t^i_j} r_{\vect{\beta}, \vect{a}^i, b^i}(b^i Y^{*ij}_s + \ell
  _{\vect{\beta},1}^{ij}(s)) ds \right] \! \! ,
\eea
where
\[
r_{\vect{\beta}, \vect{a}^i, b^i}(x) = \vect{a}^i \bar{\vect{g}}_{\vect{\beta},1}(
h^{-1}_{\vect{\beta},1}(x))^\top+ (b^i)^{-2} \vect{a}^i \tilde{\vect{g}}_{\vect{\beta},1}(
h^{-1}_{\vect{\beta},1}(x))^\top \tilde{\vect{g}}_{\vect{\beta},1}(
h^{-1}_{\vect{\beta},1}(x)) (\vect{a}^i)^\top +  (b^i)^{2}
m^{(2)}_{\vect{\beta},1}(x) 
\]
with $m^{(2)}$ given by (\ref{mmm}). 

It follows from (\ref{exponentiallikelihood1}) and standard results
on maximum likelihood estimation
for the multivariate normal and the gamma distribution that the
EM-algorithm works as follows. We use the vector and the matrix defined by
(\ref{matK}) and (\ref{matD}), but we suppress the dependence on
$\underline{\vect{a}}$ and $\underline{\vect{b}}$ in the notation and
write $\vect{v}_{\vect{\beta}}$ and $\mat{D}_{\vect{\beta}}$. 

\begin{description}
\item{0.} Choose initial values $\hat{\vect{\alpha}}_0,
  \hat{\vect{\beta}}_0, \hat{\vect{\gamma}}_0 $, $k:=0$    
\item{1.} For $i=1,\ldots,N$, generate MC-samples $\vect{X}^{m,i}_{\mbox{\tiny
      mis}} = \{ \vect{Y}^{*m,i},  \vect{a}^{m,i}, b^{m,i} \} $,
  $m=1,\ldots,M$, conditionally on $\vect{X}^i_{\mbox{\tiny obs}}$
  under the parameter values $\hat{\vect{\alpha}}_k, 
  \hat{\vect{\beta}}_k, \hat{\vect{\gamma}}_k$. Then for each $m$,
  calculate the averages 
  \bean
& \bar{\vect{a}} = \frac{1}{NM} \sum_{i,m} \vect{a}^{m,i},
\hspace{5mm}
\mat{S}= \frac{1}{NM} \sum_{i,m} (\vect{a}^{m,i} - \bar{\vect{a}})^\top (\vect{a}^{m,i}
- \bar{\vect{a}}) \\ & \\
& \bar e = \frac{1}{NM} \sum_{i,m} (b^{m,i})^{-2}, \hspace{5mm} \bar l =
-\frac{2}{NM} \sum_{i,m} \log(b^{m,i}). 
   \eean 
Finally, for each $m$ let $\vect{v}^m_{\vect{\beta}}$, $\mat{D}^m_{\vect{\beta}}$
  and $q^m(\vect{\beta})$ denote the quantities given by (\ref{matK}),
  (\ref{matD}) and (\ref{EMq})
   
 \item{2.}
Set
   \bean
   \hat{\vect{\beta}}_{k+1} &:=& \mbox{argmax}_{\vect{\beta}} \left(
     \mbox{\small $\frac12$}  \hat{\vect{v}}_{\vect{\beta}}
   \hat{\mat{D}}_{\vect{\beta}}^{-1} \hat{\vect{v}}_{\vect{\beta}}^\top +
   \hat{q}({\vect{\beta}}) \right) \\
   \hat{\vect{\alpha}}_{k+1} &:=& \hat{\vect{v}}_{\hat{\vect{\beta}}_{k+1}}
   \hat{\mat{D}}_{\hat{\vect{\beta}}_{k+1}}^{-1} \\
   \hat{\xi}_{k+1} &:=& \bar{\vect{a}}, \hspace{3mm }
   \hat{\mat{\Gamma}}_{k+1} := \mat{S}^{-1} \hspace{3mm }
   \hat{\delta}_{k+1} := \hat{\kappa}_{k+1}/\bar{e},  
   \eean
   where
\[\hat{\vect{v}}_{\vect{\beta}} = \frac1M \sum_{m=1}^M
\vect{v}^m_{\vect{\beta}}, \hspace{5mm}
\hat{\mat{D}}_{\vect{\beta}} = \frac1M \sum_{m=1}^M
\mat{D}^m_{\vect{\beta}}, \hspace{5mm}
\hat{q}({\vect{\beta}}) = \frac1M \sum_{m=1}^M
q^m({\vect{\beta}}),
\]
and where $\hat{\kappa}_{k+1}$ is the unique solution to $\log(\hat{\kappa}_{k+1})
   - \psi(\hat{\kappa}_{k+1}) = \log (\bar e) - \bar l$  ($\psi$ denotes the
   digamma function) 
\item{3.} k:=k+1; GO TO 1. 
\end{description}

The MC-samples in step 1 can be generated for each value of $i$ by
means of a simplified version of the previous Gibbs sampler in which
the following definitions are used:
\[
G_{i1} = \sum_{j=2}^{n_i}
\frac{(h_{\hat{\vect{\beta}}_k, 1}(x_j^i)-h_{\hat{\vect{\beta}}_k,
    1}(x_{j-1}^i))^2}{2(t_j^i - t_{j-1}^i)} \hspace{8mm} 
G_{i2} = - \int_{x_1^i}^{x_{n_i}^i} \frac{\hat{\vect{\alpha}}_k\vect{f}(y)^\top +
  \vect{a}^i \vect{g}(y)^\top}{c_{\hat{\vect{\beta}}_k}^2(y)}dy 
\]
and 
\[
F_i(b) = 
- \frac12 \sum_{j=2}^{n_i}
\int_{t_{j-1}^i}^{t_j^i} \phi_{\hat{\vect{\alpha}}_k,
  \hat{\vect{\beta}}_k, \vect{a}^i, b}(Y^{*ij}_s +
  b^{-1}\ell_{\hat{\vect{\beta}}_k, 1}^{ij}(s)) ds,
\]
where
\[
\phi_{\hat{\vect{\alpha}}_k,  \hat{\vect{\beta}}_k, \vect{a}^i, b}(y)
= m_{\hat{\vect{\alpha}}_k,  \hat{\vect{\beta}}_k, \vect{a}^i,
  1}^{(1)}(b y) + b^2 m_{\hat{\vect{\beta}}_k, 1}^{(2)}(b y) + b^{-2}
m_{\hat{\vect{\alpha}}_k,  \hat{\vect{\beta}}_k, \vect{a}^i, 1}^{(3)}(b y)  
\]
with $m^{(1)}, m^{(2)}$ and $m^{(3)}$ given by (\ref{mmm}). 

The Gibbs sampler goes as follows, cf.\ (\ref{exponentiallikelihood2}).
\begin{description}

\item{0.} Draw independent values of $\vect{a}^i$ and $E$ from the
  $N_{p_2}(\hat{\vect{\xi}}_k,
  \hat{\mat{\Gamma}}_k^{-1})$-distribution and the $\Gamma( \hat
  \kappa_k, \hat \delta_k)$-distribution, respectively, and set
  $b^i := E^{-2}$ 

\item{1.} Simulate independent sample paths $Y^{*ij}$, $j=2, \ldots,
  n_i$, conditionally on $\vect{a}^i, b^i,
  \vect{X}^i_{\mbox{\tiny obs}}$ (with the parameter values
  $\hat{\vect{\alpha}}_k$ and $\hat{\vect{\beta}}_k$), and use these
  to calculate $\vect{t}_{\hat{\vect{\alpha}}_k, \hat{\vect{\beta}}_k,b^i}^i$
  and $\mat{B}_{\hat{\vect{\beta}}_k, b^i}^i$
  given by (\ref{matA}) and (\ref{matB}). 

\item{2.} Draw $\vect{a}^i$ from the $N_{p_2} (
  (\vect{t}_{\hat{\vect{\alpha}}_k, \hat{\vect{\beta}}_k, b^i}^i
  +\hat{\vect{\xi}}_k  \hat{\mat{\Gamma}}_k)(\mat{B}_{\hat{\vect{\beta}}_k,
    b^i}^i + \hat{\mat{\Gamma}}_k)^{-1}, 
  (\mat{B}_{\hat{\vect{\beta}}_k, b^i}^i
  +\hat{\mat{\Gamma}}_k)^{-1})$-distribution (conditionally on $b^i, 
  \vect{X}^i_{\mbox{\tiny obs}}$ and $\vect{Y}^{*i}$) 

\item{3.} Draw $E$ from the weighted gamma distribution with density
  proportional to
\be
\label{weightedgammab}
x^{(n_i-1)/2 + \hat \kappa_k -1}\exp\left( - x (\hat \delta_k + G_{i1} + G_{i2})
  + F_i(x^{-1/2})  \right), \ \ x > 0
\ee
 (conditionally on $a^i, \vect{X}^i_{\mbox{\tiny obs}}$ and
 $\vect{Y}^{*i}$), and set $b^i := E^{-2}$ 

\item{4.} GO TO 1
\end{description}

Draws from the weighted gamma distribution (\ref{weightedgammab}) can
be done in various ways, as indicated in the previous subsection. Here
we just mention that if $\hat{\delta}_k + G_{i1} + G_{i2} > 0$, then
step 3 in the Gibbs sampler can be replaced by
\begin{description}
\item{3*.} Draw $E$ from the $\Gamma ( (n_i - 1)/2 + \hat{\kappa}_k,  \hat{\delta}_k +
G_{i1} + G_{i2})$-distribution. With probability
\[
\min \left( 1, \exp(F_i(E^{-1/2}) - F_i(b^i)) \right)
\]
the proposed value is accepted and $b^i := E^{-2}$. Otherwise $b^i$
is unchanged.
\end{description}
If $\hat{\delta}_k + G_{i1} + G_{i2} \leq 0$, the scale parameter of
the gamma distribution must be changed to $(\hat{\delta}_k +
G_{i1})^{-1}$, and $F_i$ must be replaced by
the function $\tilde F_i (b) = b^{-2} G_{i2} + F_i(b)$. 

If some of the random effects in the drift must necessarily be
positive, these coordinates of the multivariate normal distribution
can restricted to the positive half-line. Perhaps a more satisfactory
solution is to assume that these coordinates are exponential
distributed and independent of
the other coordinates. For instance, if $a^i$ is one-dimensional
($p_2=1$) and must be positive, then we can assume that $a^i$ is
exponential distributed with mean $\lambda^{-1}$. Then in step 2 of
the EM-algorithm $\hat \xi_{k+1}$ and $\hat{\mat{\Gamma}}_{k+1}$ are
replaced by 
\[
\hat \lambda_{k+1}:= 1/\bar a
\]
(and $\mat{S}$ is not needed in step 1). In step 2 of the Gibbs sampler,
$a^i$ is drawn from the normal distribution with mean
$(t_{\hat{\vect{\alpha}}_k, \hat{\vect{\beta}}_k, b^i}^i - \hat 
\lambda_k)/B_{\vect{\beta}, b^i}^i$ and variance
$(B_{\vect{\beta}, b^i}^i)^{-1}$ restricted to the positive 
half-axis (conditionally on $b^i, \vect{X}^i_{\mbox{\tiny obs}}$ and
$\vect{Y}^{*i}$). 

\sect{Simulation studies}
\label{simul}

\subsection{The Ornstein-Uhlenbeck process}

 Consider the Ornstein-Uhlenbeck process 
\[
dX^i_t = -a^iX^i_tdt + \beta dW^i_t,
\]
with random speed parameter $a^i$, where $\beta > 0$ and $a^i$ is
exponential distributed with mean $\gamma^{-1}$. 

First we apply the {\it Gibbs sampler}. As the prior, we choose the
$\Gamma(\nu, \lambda)$ distribution for $\gamma$ and the
$\Gamma(\kappa, \delta)$ distribution for $\eta = \beta^{-2}$, and we
assume independence of $\gamma$ and $\beta$.

Since $f(x)=0, g(x) = -x$, $h_\beta(x) = x/\beta$ and $\phi_a(x) = -a
+ a^2x^2$,  the quantities needed in the algorithm are
\bean
t^i_\beta &=& -\frac{1}{2\beta^2} \left( (x_{n_i}^i)^2 - (x_1^i)^2 \right) +
\frac12 (t^i_{n_i} - t^i_1) \\
B^i_\beta &=& \sum_{j=2}^{n_i} \int_{t^i_{j-1}}^{t^i_j} \left(  Y^{*ij}_s +
  \beta^{-1}\ell_1^{ij}(s)\right)^2 ds \\
G_1 &=& \sum_{i=1}^N \sum_{j=2}^{n_i} \frac{(x_j^i -
  x_{j-1}^i)^2}{2(t_j^i-t_{j-1}^i)} \\
G_2 &=& \frac12 \sum_{i=1}^N a^i \left( (x_{n_i}^i)^2 -
  (x_1^i)^2\right) \\
E_1 &=& \frac12 \sum_{i=1}^N (a^i)^2 \sum_{j=2}^{n_i}
\int_{t^i_{j-1}}^{t_j^i} \ell^{ij}_1(s)^2 ds \\
E_2 &=& - \sum_{i=1}^N  (a^i)^2 \sum_{j=2}^{n_i}
\int_{t^i_{j-1}}^{t_j^i} Y_s^{*ij} \ell^{ij}_1(s) ds,
\eean
where
\be
\label{ell}
\ell^{ij}_1(t) =
\frac{(t^i_j-t)x^i_{j-1}
+ (t-t^i_{j-1})x^i_j}
{t^i_j-t^i_{i-j}}, \hspace{5mm} t \in  [t^i_{j-1}, t^i_j].
\ee
In this example $F(\beta) = -\beta^{-2} E_1 + \beta^{-1} E_2$ (apart
from an additive term independent of $\beta$), but here we can make $E_1$
a part of the scale parameter of the weighted gamma distribution, because it
is positive.

The Gibbs sampler goes as follows, with
$N_+$ denoting the normal distribution restricted to the positive 
half-line.

\begin{description}

\item{0.} First draw $\beta$ and $\gamma$ independently from the prior
  distribution, and given $\gamma$, draw $a^i$ from the exponential
  distribution with mean $\gamma^{-1}$, independently for
  $i=1,\ldots,N$.   
  
\item{1.} Simulate independent sample paths $Y^{*ij}$ conditionally on
  $a^i, \beta$ and $X^i_{\mbox{\tiny obs}}$ for $j=2, \ldots, n_i, \ i=1, \ldots, N$

\item{2.} Draw $a^i$ with distribution $N_+((t_\beta^i - \gamma)/B^i_\beta,
  (B^i_\beta)^{-1})$, independently for $i=1,\ldots,N$  

\item{3.} Draw $\eta$ from the distribution with
  density function proportional to
  \[
\eta^{(n_\cdot-N)/2 + \kappa -1}\exp \left( - \eta (\delta + G_1 + G_2 +
E_1) + \sqrt{\eta} E_2 \right), 
    \]
and set $\beta := \eta^{-1/2}$

\item{4.} Draw $\gamma$ from the $\Gamma ( \nu +n, \lambda + a^1 +
  \cdots + a^N)$-distribution

\item{5.} GO TO 1

\end{description}

\noindent
If $\delta + G_1 + G_2 + E_1>0$, the distribution in step 3 is a
weighted gamma distribution with scale parameter $(\delta + G_1 + G_2
+ E_1)^{-1}$ and weight function $\exp (\sqrt{\eta} E_2)$.
If $\delta + G_1 + G_2 + E_1 \leq 0$, the scale parameter is
$(\delta + G_1 + E_1)^{-1}$ (which is always positive) and the weight
function is $\exp (\sqrt{\eta} E_2 - \eta G_2)$.

As data we simulated 100 diffusions with $\beta = 1$ and $\gamma = 1$ at the
time points $t^i_j = j, j=1, \ldots, 100$. Then we ran 1000 iterations of
the Gibbs sampler using prior distributions with $\kappa = 1, \delta =
0.5, \nu = 1, \lambda = 2$. In step 1 we simulated the approximate
diffusion bridges proposed by \cite{bladtsorensen}, and in step 3 we used the
approximate direct sampling method described in Subsection
\ref{expGibbs}. In all cases $\delta + G_1 + G_2 + E_1$ was positive. 

\begin{figure}
\begin{center}
\includegraphics[width=4.0cm]{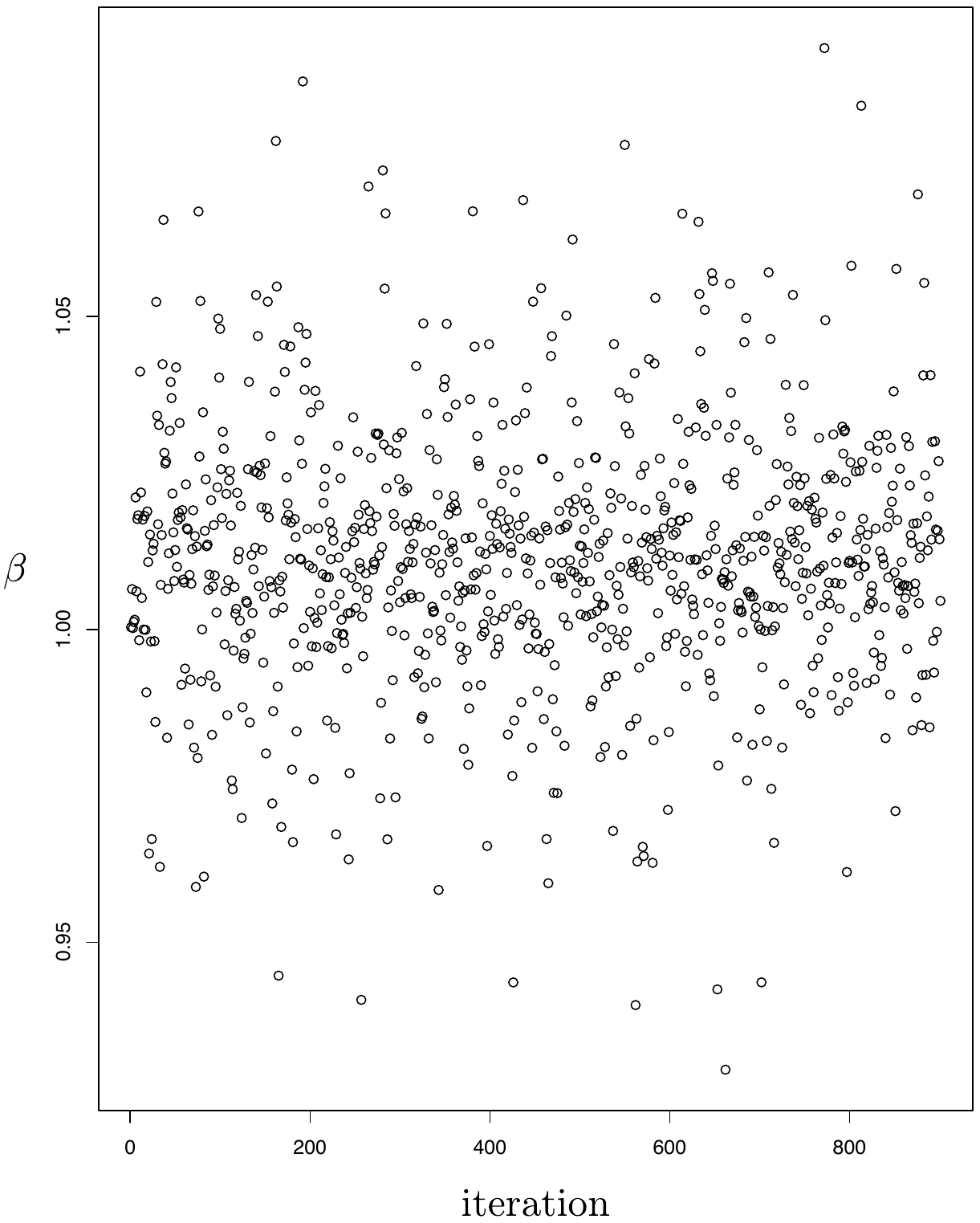} 
\includegraphics[width=4.0cm]{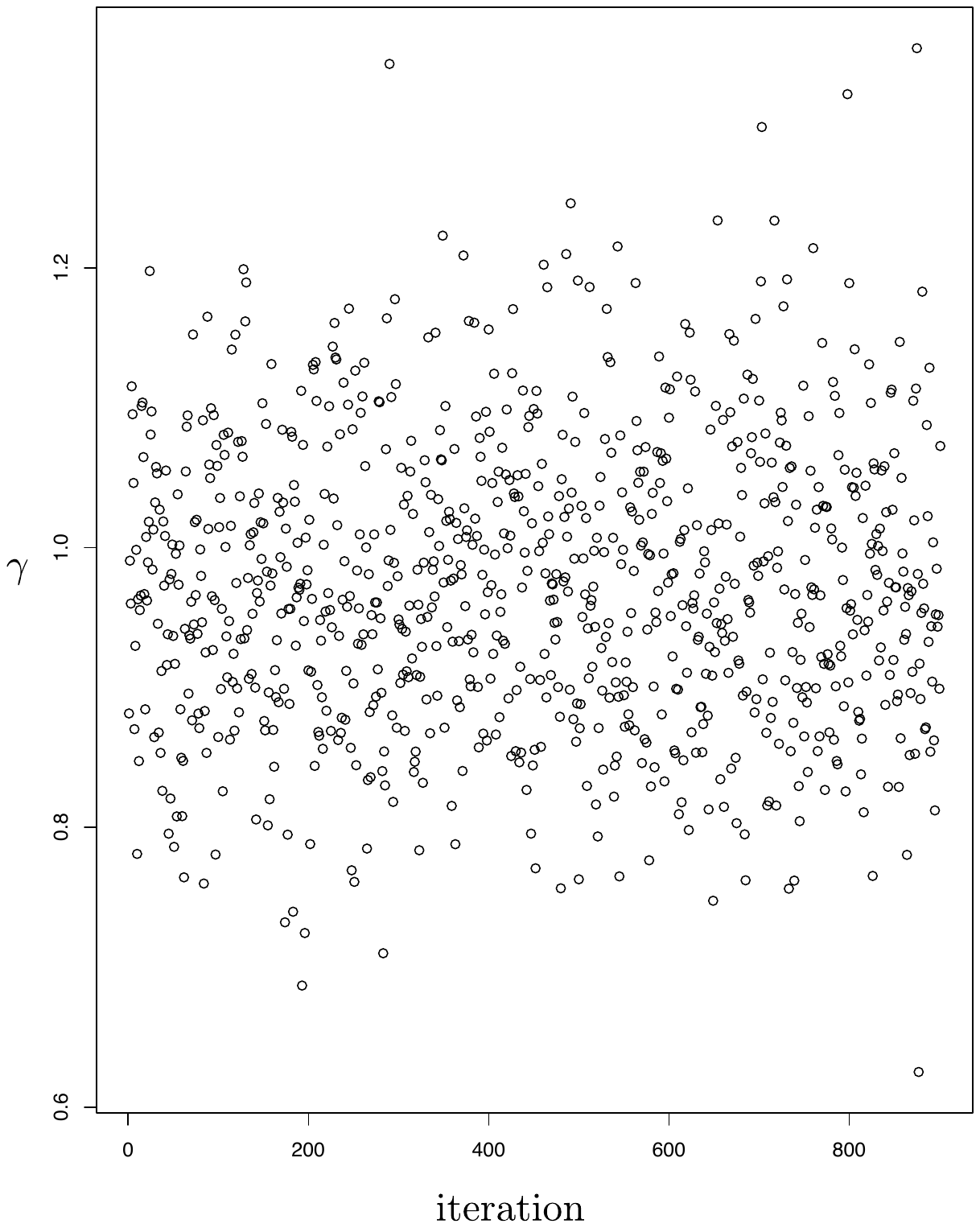}
\includegraphics[width=4.0cm]{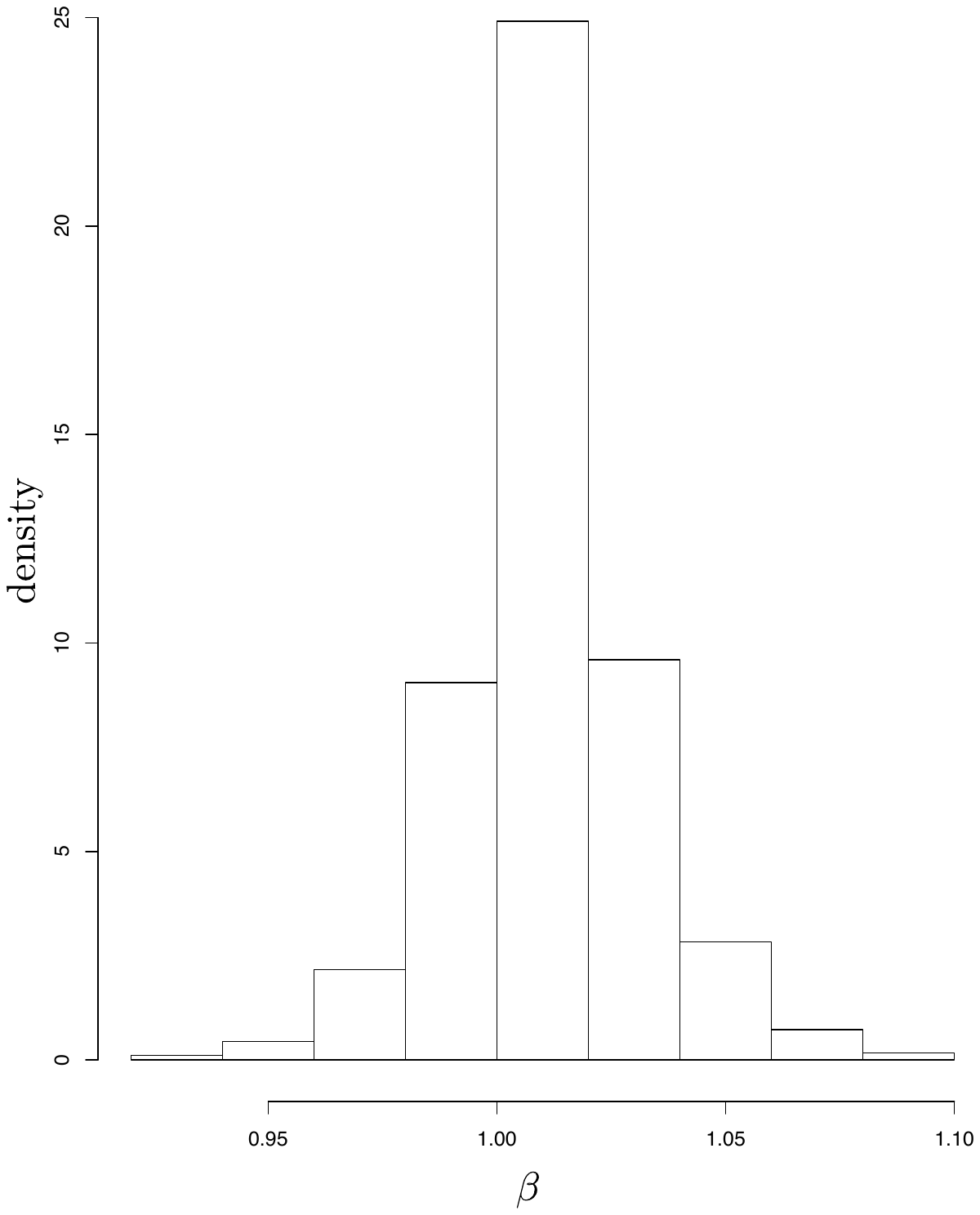}
\includegraphics[width=4.0cm]{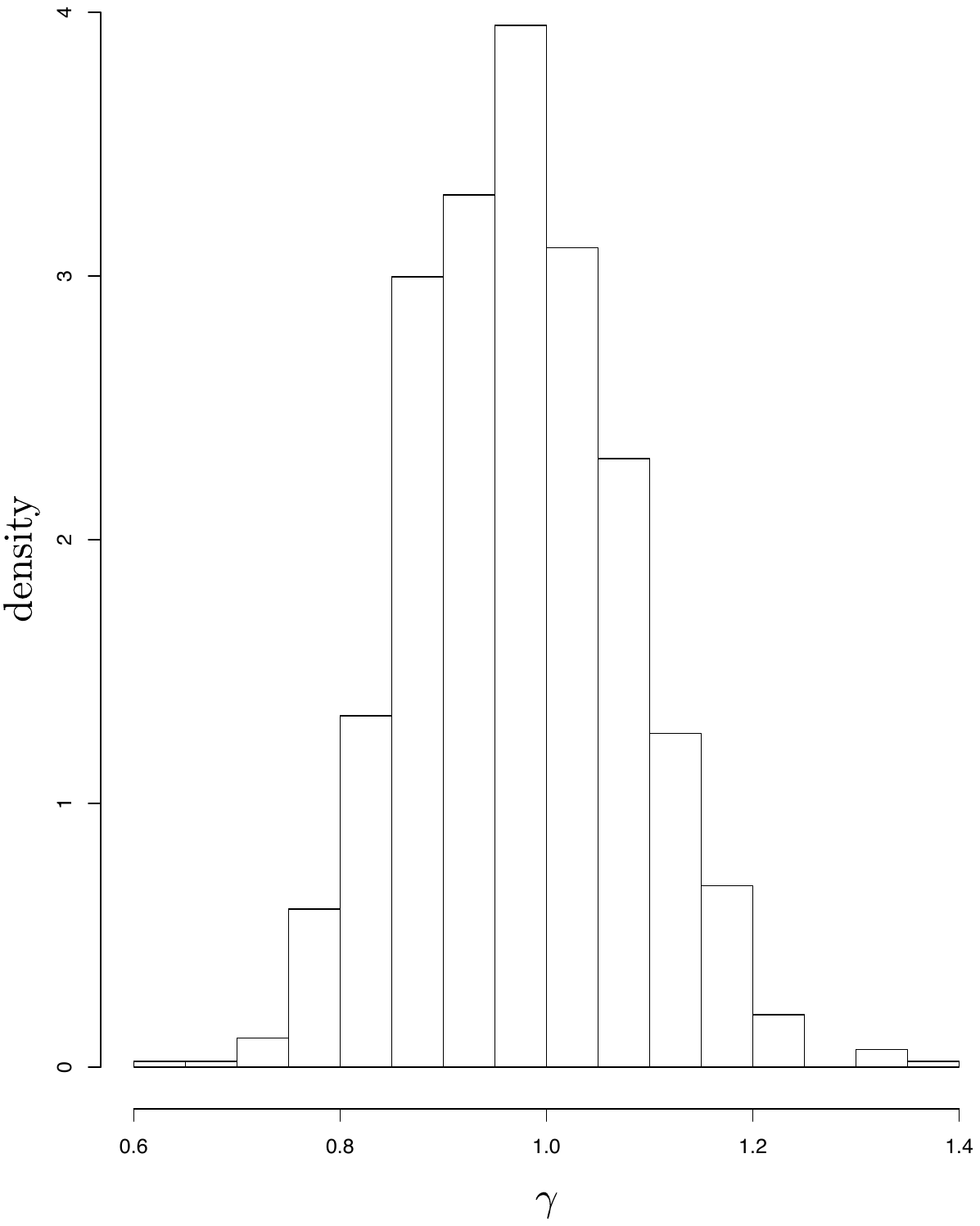}
\caption{\label{simulationOU1} 
The Gibbs sampler for the Ornstein-Uhlenbeck process with random drift
parameter: scatter plots of the draws of the parameters $\beta$ and
$\gamma$ (after a burn-in of 100 draws) and histograms of the
draws. The true parameter values are $\beta = 1$ and  $\gamma =1$.} 
\end{center}
\end{figure}

Because the burn-in was almost immediate, we based the estimation on the last
900 draws of the parameters. The mean posterior estimators are
$\hat{\beta} =0.9800$ and $\hat{\gamma}=1.017$, and the $95\%$
credibility intervals are $[ 0.9368, 1.0200 ]$ for $\beta$ and $[
0.8264, 1.2291 ]$ for $\gamma$.  The last 900 draws and histograms of
the parameter values  are plotted in Figure \ref{simulationOU1}.

Next we apply the {\it EM algorithm}. In this example
\[
q(\beta) = -\beta^{-2}(G_1 + E_1) + \beta^{-1}E_2
-(n_\cdot - N)\log(\beta)
\]
(apart from additive terms that do not depend on $\beta$). The
EM-algorithm goes as follows:

\begin{description}
\item{0.} Choose initial values $\hat{\beta}_0$ and $\hat{\gamma}_0$, $k:=0$    
\item{1.} For $i=1,\ldots,N$, generate MC-samples $\vect{X}^{m,i}_{\mbox{\tiny
      mis}} = \{ \vect{Y}^{*m,i},  a^{m,i} \} $,  $m=1,\ldots,M,$
  conditionally on $\vect{X}^i_{\mbox{\tiny obs}}$ under the parameter
  values  $\hat{\beta}_k$ and $\hat{\gamma}_k$,  and for each $m$
  calculate $G_2^m, E_1^m$ and $E_2^m$. Finally, calculate the averages 
 \[
   \hat{G}_2 = \frac1M \sum_{m=1}^M G^m_2, \hspace{6mm}
   \hat{E}_1= \frac1M \sum_{m=1}^M E^m_1, \hspace{6mm}
   \hat{E}_2 = \frac1M \sum_{m=1}^M E^m_2, \hspace{6mm}
   \bar{a} = \frac{1}{NM} \sum_{i,m} a^{m,i}
\]
 \item{2.}
\[
   \hat{\beta}_{k+1} := \mbox{argmax}_{\vect{\beta}} \left(
     -\beta^{-2}(G_1 + \hat{G}_2 + \hat{E}_1) + \beta^{-1}\hat{E}_2
     -(n_\cdot - N)\log(\beta) \right), \hspace{7mm}
   \hat{\gamma}_{k+1} := 1/\bar{a}
\]
\item{3.} k:=k+1; GO TO 1. 
\end{description}
The maximization in step 2 is elementary. If $G_1 + \hat{G}_2  + \hat{E}_1>0$ (which
is usually the case), then  
\[
\hat{\beta}_{k+1} = \frac{4(G_1 + \hat{G}_2  + \hat{E}_1)}{-\hat{E}_2 +
  \sqrt{\hat{E}^2_2 + 8 (G_1 + \hat{G}_2  + \hat{E}_1)(n_\cdot - N)}}.
\]
Moreover, $\hat{\beta}_{k+1} = - 4(G_1  + \hat{G}_2  + \hat{E}_1)/(\hat{E}_2 +
\sqrt{\hat{E}^2_2 + 8 (G_1 + \hat{G}_2  + \hat{E}_1)(n_\cdot - N)}$ when $0 > G_1 +
\hat{G}_2 + \hat{E}_1 \geq - \frac18 \hat{E}^2_2/(n_\cdot - N)$ and $\hat{E}_2>0$,
and if $G_1 + \hat{G}_2 + \hat{E}_1=0$ and $\hat{E}_2<0$, then
$\hat{\beta}_{k+1} = - \hat{E}_2 /(n_\cdot - N)$. In other cases, a
positive maximum does not exist.
  
The MC-samples in step 1 can be generated for each value of $i$ by
means of the following simplified version of the previous Gibbs sampler.
\begin{description}
\item{0.} Draw $a^i$ from the exponential distribution with mean
  $\hat{\gamma}_k^{-1}$
\item{1.} Simulate independent sample paths $Y^{*ij}$, $j=2, \ldots,
  n_i$, conditionally on $a^i$ and $\vect{X}^i_{\mbox{\tiny obs}}$
  (with the parameter value $\hat{\beta}_k$), and use these
  to calculate $t^i_{\hat{\beta}_k}$ and $B^i_{\hat{\beta}_k}$ 

\item{2.} Draw $a^i$ from the $N_{+} ((t^i_{\hat{\beta}_k} -
  \hat{\gamma}_k)/B^i_{\hat{\beta}_k}, (B^i_{\hat{\beta}_k})^{-1})$-distribution

\item{3.} GO TO 1
\end{description}

\begin{figure}
\begin{center}
\includegraphics[width=7.0cm]{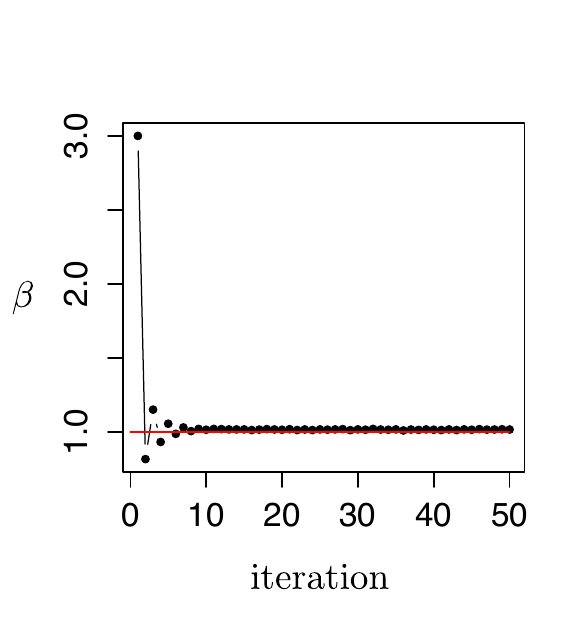} 
\includegraphics[width=7.0cm]{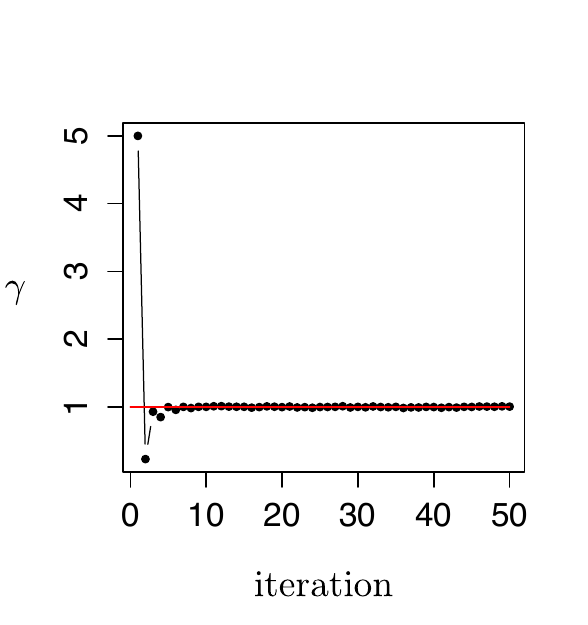}
\caption{\label{simulationOU2} 
The EM algorithm for the Ornstein-Uhlenbeck process with random drift
parameter: parameter values in 50 iterations. The true parameter
values are $\beta = 1$ and  $\gamma =1$.} 
\end{center}
\end{figure}

We simulated 100 diffusions with $\beta = 1$ and $\gamma = 1$ at the
time points $t^i_j = j, j=1, \ldots, 100$ as our data. Then we ran 50 iterations of
the EM algorithm with initial values $\hat{\beta}_0 = 3$ and
$\hat{\gamma}_0 = 5$.
The parameter values in the iterations are plotted in Figure
\ref{simulationOU2}. The convergence was fast. As
estimators we use the final parameter values. The EM maximum likelihood
estimators  estimators are $\hat{\beta} = 1.017$ and $\hat{\gamma}=1.003$.

\subsection{A t-diffusion}

Consider the $t$-diffusion 
\[
dX^i_t = -a^i X^i_t dt + \beta \sqrt{1+(X^i_t)^2} dW^i_t,
\]
with random speed parameter $a^i$, where $\beta > 0$ and $a^i$ is
exponential distributed with mean 
$\gamma^{-1}$.  The t-diffusion is one of the Pearson diffusions; see
\cite{forman}. For fixed $a^i$ it is an ergodic process on $\R$
provided that $a^i > 0$. The invariant probability measure is a
re-scaled $t$-distribution with $\nu = 2 a^i/\beta^2+1$
degrees of freedom (and scale-parameter $\nu^{-\frac12}$).

For this model
\bean
h_\beta(x) &=& \beta^{-1} \log(x+\sqrt{x^2+1}) \\
h_\beta^{-1}(y) &=& \sinh (\beta y) \\
\mu_{a^i,\beta}(y) &=& -(a^i/\beta+\beta/2) \tanh (\beta y) \\
\phi_{a^i,\beta}(y) &=& (a^i/\beta+\beta/2)^2 
\tanh^2 (\beta y) - \frac{a^i + \frac12 \beta^2}{\cosh^2(\beta
  y)} \\
s_{a^i,\beta}(x) &=& -\frac{a^i}{2 \beta^2} \log (1+x^2).
\eean
Note that $a^i/\beta+\beta/2 = \frac12 \beta \nu$.

First we consider the {\it Gibbs sampler}. As the prior, we choose
again the $\Gamma(\nu, \lambda)$ distribution for $\gamma$ and the
$\Gamma(\kappa, \delta)$ distribution for $\eta = \beta^{-2}$, and we
assume independence of $\gamma$ and $\beta$. In the algorithm we use
following quantities
\bean
t^i_\beta &=& - \frac{1}{2 \beta^2} \log \left( \frac{1+
    (x^i_{n_i})^2}{1+ (x^i_1)^2} \right) +\frac12 (t^i_{n_i} - t^i_1)
- \sum_{j=2}^{n_i} \int_{t^i_{j-1}}^{t^i_j} \tanh^2 (\beta Y^{*ij}_s +
\ell_1^{ij}(s)) ds \\
B^i_\beta &=& \frac{1}{\beta^2} \sum_{j=2}^{n_i}
\int_{t^i_{j-1}}^{t^i_j} \tanh^2 (\beta Y^{*ij}_s + \ell_1^{ij}(s)) ds \\
G_1 &=& \sum_{i=1}^N \sum_{j=2}^{n_i} \frac{\left( \log \left(
      \frac{x^i_{j} + \sqrt{(x^i_{j})^2 +1}}{x^i_{j-1} +
        \sqrt{(x^i_{j-1})^2 +1}} \right)\right)^2}{2(t^i_{j-1}
  -t^i_j)} \\
G_2 &=& \frac12 \sum_{i=1}^N a^i \log \left(
  \frac{1+(x^i_{n_i})^2}{1+(x^i_1)^2} \right)  \\
F(\beta) &=& -\frac12 \sum_{i=1}^N \left[ \left( \frac{(a^i)^2}{\beta^2} +
  \frac34 \beta^2 +a^i (\beta +1) \right)  \sum_{j=2}^{n_i}
\int_{t^i_{j-1}}^{t^i_j} \tanh^2 (\beta Y^{*ij}_s + \ell_1^{ij}(s)) ds
\right. \\
&& \left. \hspace{9cm}- \left(  a^i +\frac12 \beta^2\right) (t^i_{n_i} - t^i_1) \right]
\eean
where
\[
  \ell_1^{ij}(t) = \frac{(t^i_j-t)
    \log\left(x^i_{j-1} + \sqrt{(x^i_{j-1})^2 +1} \, \right)
+ (t-t^i_{j-1}) \log\left(x^i_{j} + \sqrt{(x^i_{j})^2 +1} \, \right)}
{t^i_j-t^i_{i-j}}, \hspace{3mm} t \in  [t^i_{j-1}, t^i_j].
\]
With these definitions the Gibbs sampler is as in the previous example
except that step 3 is replaced by step 2* in Subsection \ref{expGibbs}. 

As data we simulated 100 diffusions with $\beta = 0.1$ and $\gamma = 1$ at the
time points $t^i_j = j, j=0, \ldots, 100$. We then ran 1000 iterations of
the Gibbs sampler using prior distributions with $\kappa = 1, \delta =
5, \nu = 1, \lambda = 0.75$. We used the approximate diffusion bridges
of \cite{bladtsorensen}. In all cases it turned out that $\delta + G_1 + G_2>0$.
 
\begin{figure}
\begin{center}
\includegraphics[width=4.0cm]{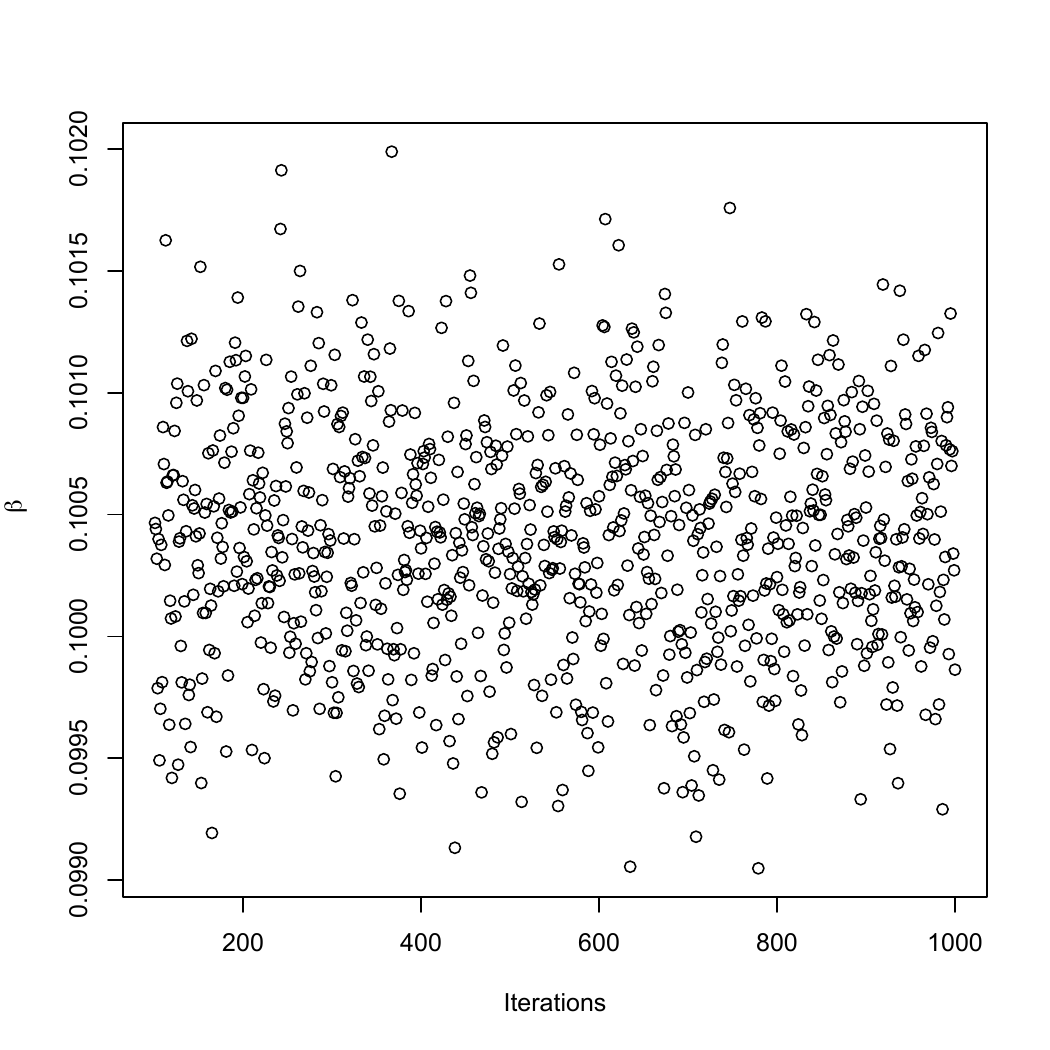} 
\includegraphics[width=4.0cm]{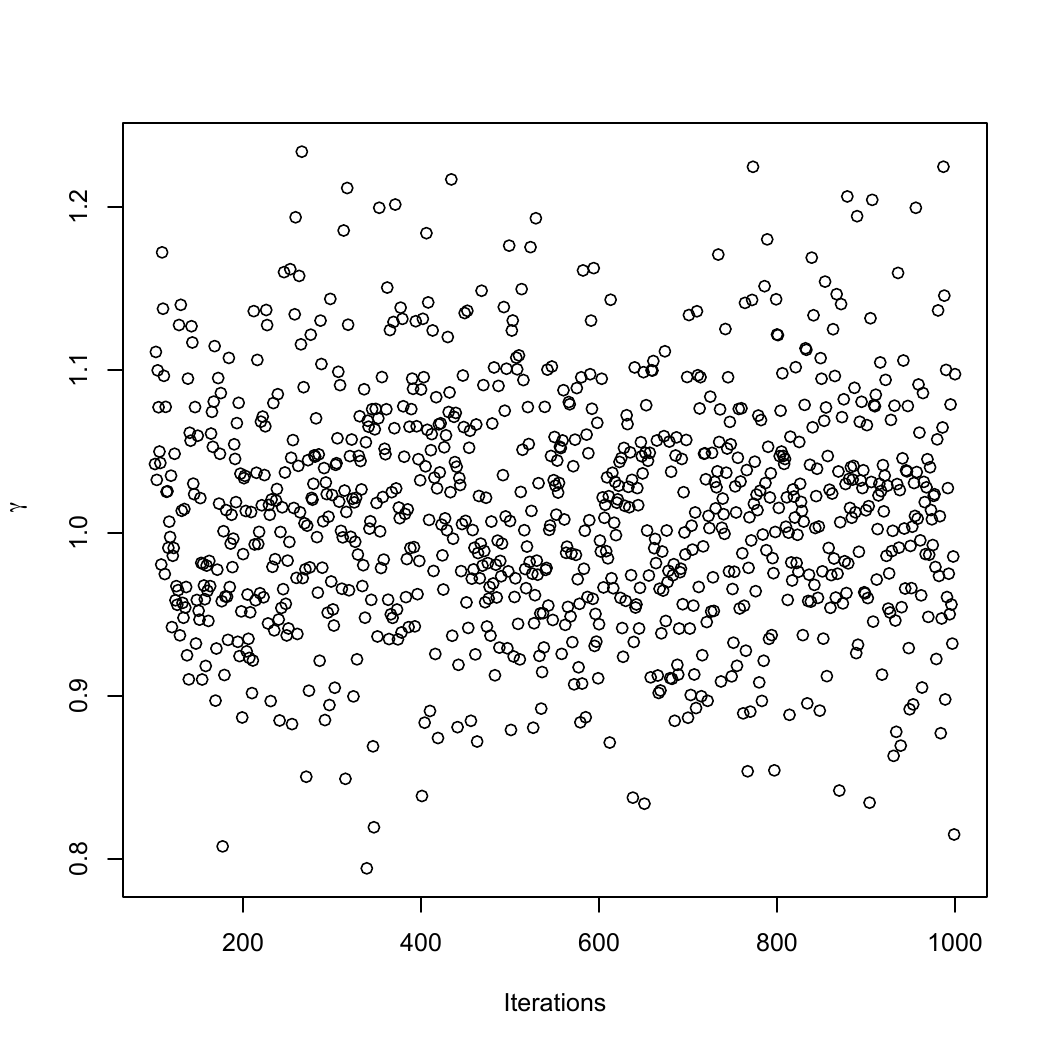}
\includegraphics[width=4.0cm]{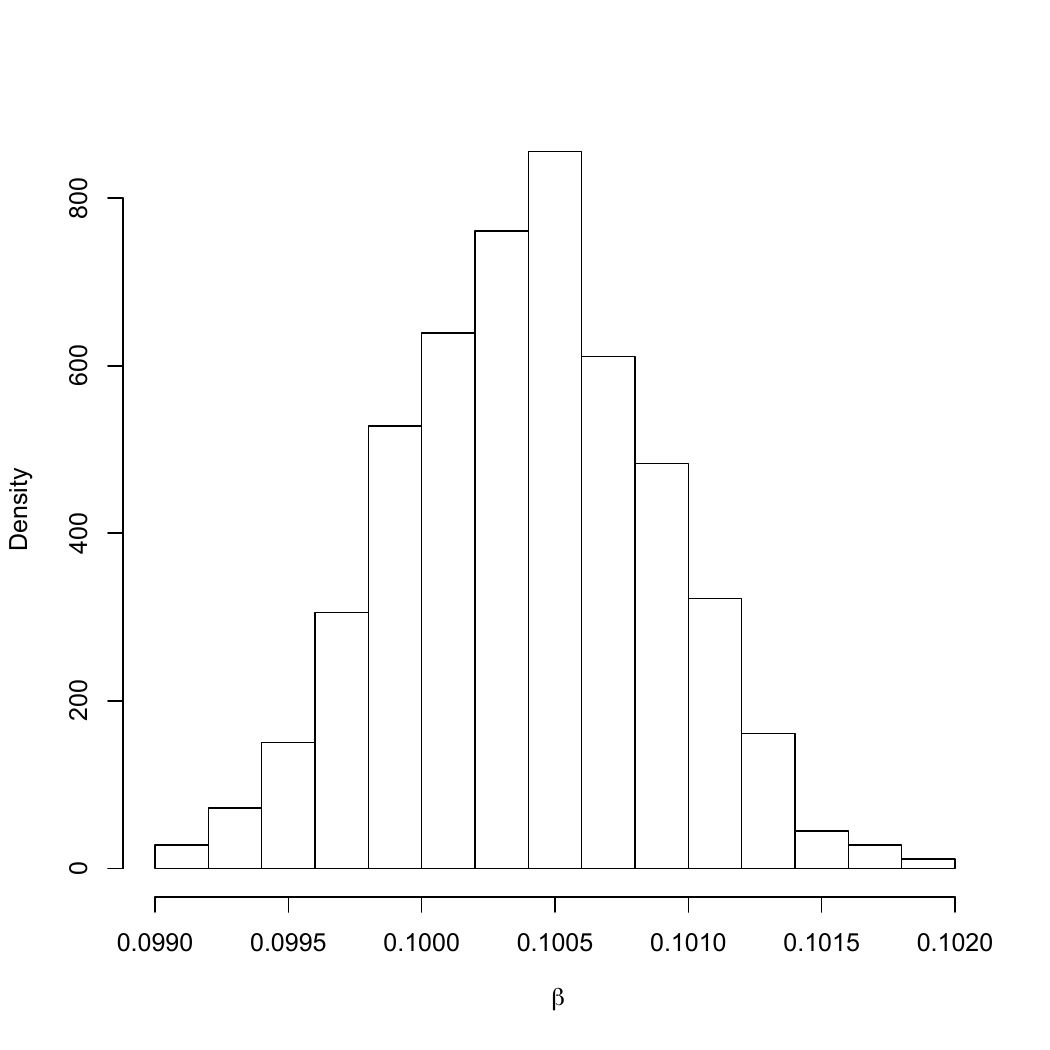}
\includegraphics[width=4.0cm]{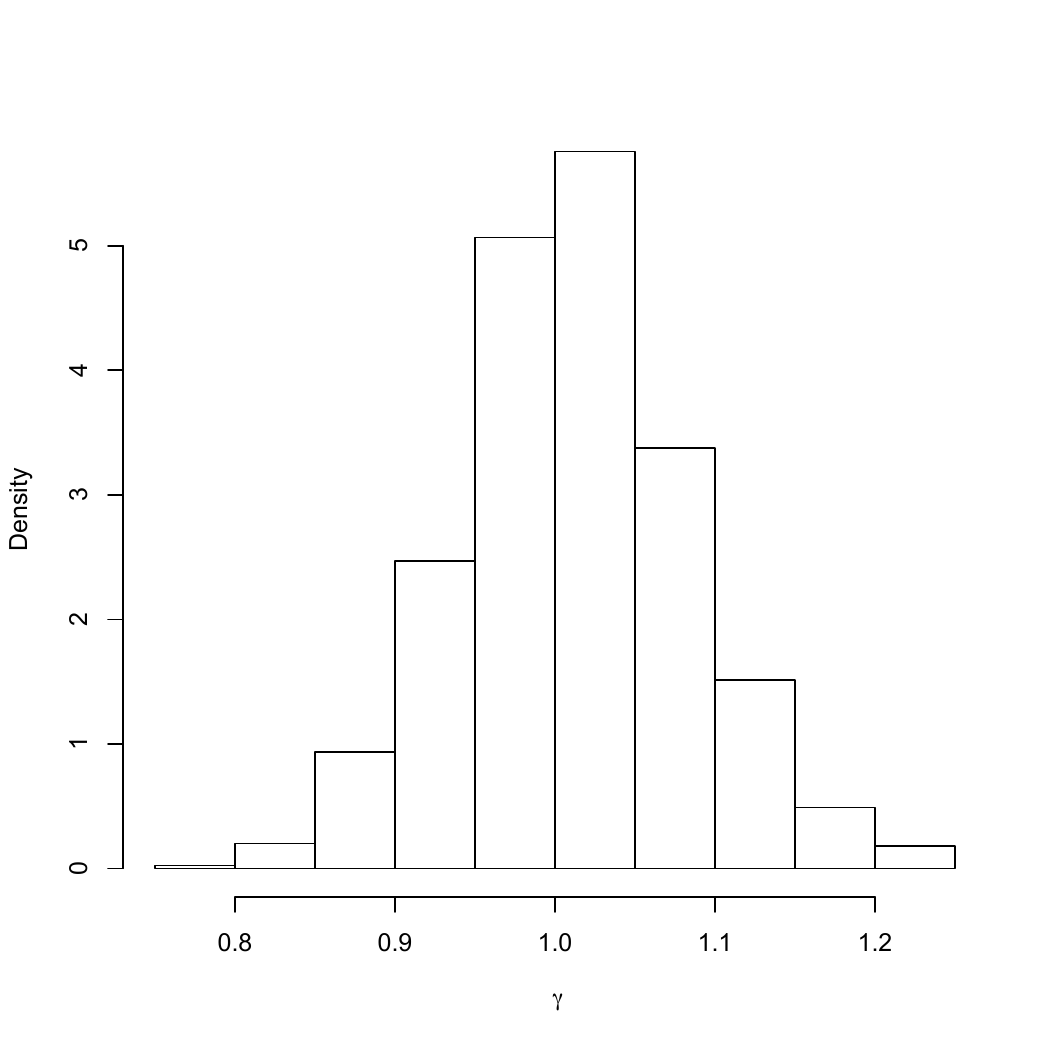}
\caption{\label{simulation_t} 
The Gibbs sampler for the t-diffusion with random drift
parameter: scatter plots of the draws of the parameters $\beta$ and
$\gamma$ (after a burn-in of 100 draws) and histograms of the
draws. The true parameter values are $\beta = 0.1$ and  $\gamma =1$.} 
\end{center}
\end{figure}

The estimation is based on the last 900 draws of the parameters, after
a burn-in of 100 iterations. The mean posterior estimators are
$\hat{\beta} =0.0998$ and $\hat{\gamma}=1.0248$, and the $95\%$
credibility intervals are $[ 0.0989, 0.1008 ]$ for $\beta$ and $[
0.8944, 1.1648 ]$ for $\gamma$.  The last 900 draws and histograms of
the parameter values  are plotted in Figure \ref{simulation_t}.

Next we turn to the {\it EM algorithm}. For the t-diffusion
\bean
q(\beta) &=& -\beta^{-2}(G_1 + G_2) + \mbox{\small $\frac14$} 
\beta^2 \sum_{i=1}^N (t^i_{n_i} - t^i_1) -(n_\cdot - N)\log(\beta) \\
&& - \mbox{\small $\frac12$} \sum_{i=1}^N \left( 2 a_i + a_i^2\beta^{-2} +
  \mbox{\small $\frac34$}  \beta^2 \right) \sum_{j=2}^{n_i}
\int_{t^i_{j-1}}^{t^i_j} \tanh^2 (\beta Y^{*ij}_s +
\ell_1^{ij}(s)) ds
\eean
(apart from additive terms that do not depend on $\beta$). The
EM-algorithm goes as follows:

\begin{description}
\item{0.} Choose initial values $\hat{\beta}_0$ and $\hat{\gamma}_0$, $k:=0$    
\item{1.} For  $i=1,\ldots,N$, generate MC-samples $\vect{X}^{m,i}_{\mbox{\tiny
      mis}} = \{ \vect{Y}^{*m,i},  a^{m,i} \} $,  $m=1,\ldots,M$ conditionally on 
$\vect{X}^i_{\mbox{\tiny obs}}$ under the parameter values
$\hat{\beta}_k$ and $\hat{\gamma}_k$, and for each $m$ calculate
$G_2^m$. Finally, calculate 
 \[
 \hat{G}_2 = \frac1M \sum_{m=1}^M G^m_2, \hspace{7mm}
   \bar{a} = \frac{1}{MN} \sum_{i,m} a^{m,i} 
\]
 \item{2.}
\vspace{-4mm}
\bean
\hat{\gamma}_{k+1} &:=& 1/\bar{\vect{a}} \\
   \hat{\beta}_{k+1} &:=& \mbox{argmax}_{\vect{\beta}} \left(
     -\beta^{-2}(G_1 + \hat{G}_2) + \mbox{\small $\frac14$} 
\beta^2 \sum_{i=1}^N (t^i_{n_i} - t^i_1) -(n_\cdot - N)\log(\beta)
- \tilde{q}(\beta) \right),   
\eean
where
\[
\tilde{q} (\beta) = \mbox{\small $\frac12$} \frac{1}{M} \sum_{i,m}
\left( 2 a^m_i + (a^m_i)^2\beta^{-2} +
  \mbox{\small $\frac34$}  \beta^2 \right) \sum_{j=2}^{n_i}
\int_{t^i_{j-1}}^{t^i_j} \tanh^2 (\beta Y^{*m,ij}_s + \ell_1^{ij}(s)) ds
\]
\item{3.} k:=k+1; GO TO 1. 
\end{description}

As in the previous example, the MC-samples in step 1 can be generated
for each value of $i$ by means of a simplified version of the previous
Gibbs sampler, which is identical to the previous sampler (except that
$t^i_{\hat{\beta}_k}$ and $B^i_{\hat \beta_k}$ are defined as in this subsection).

\begin{figure}
\begin{center}
\includegraphics[width=7.0cm]{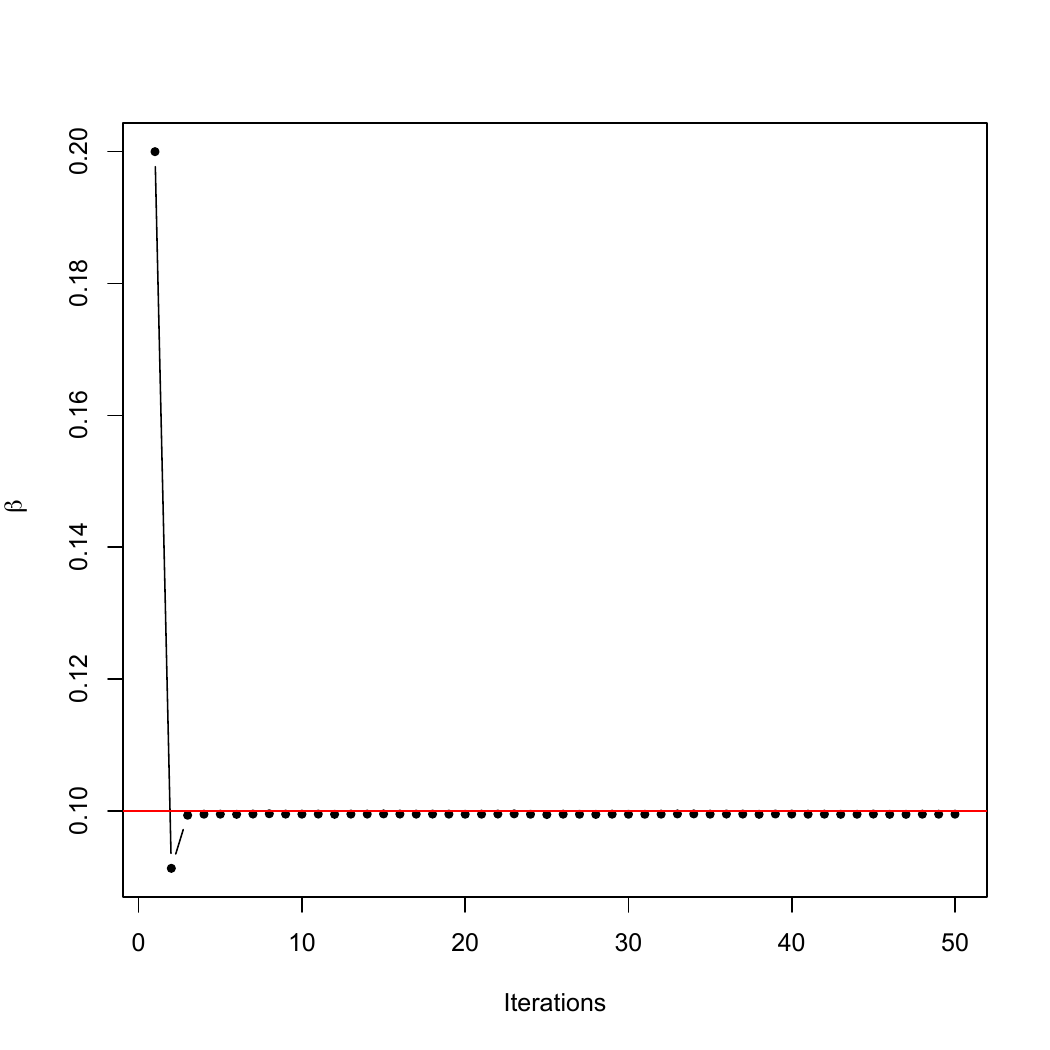} 
\includegraphics[width=7.0cm]{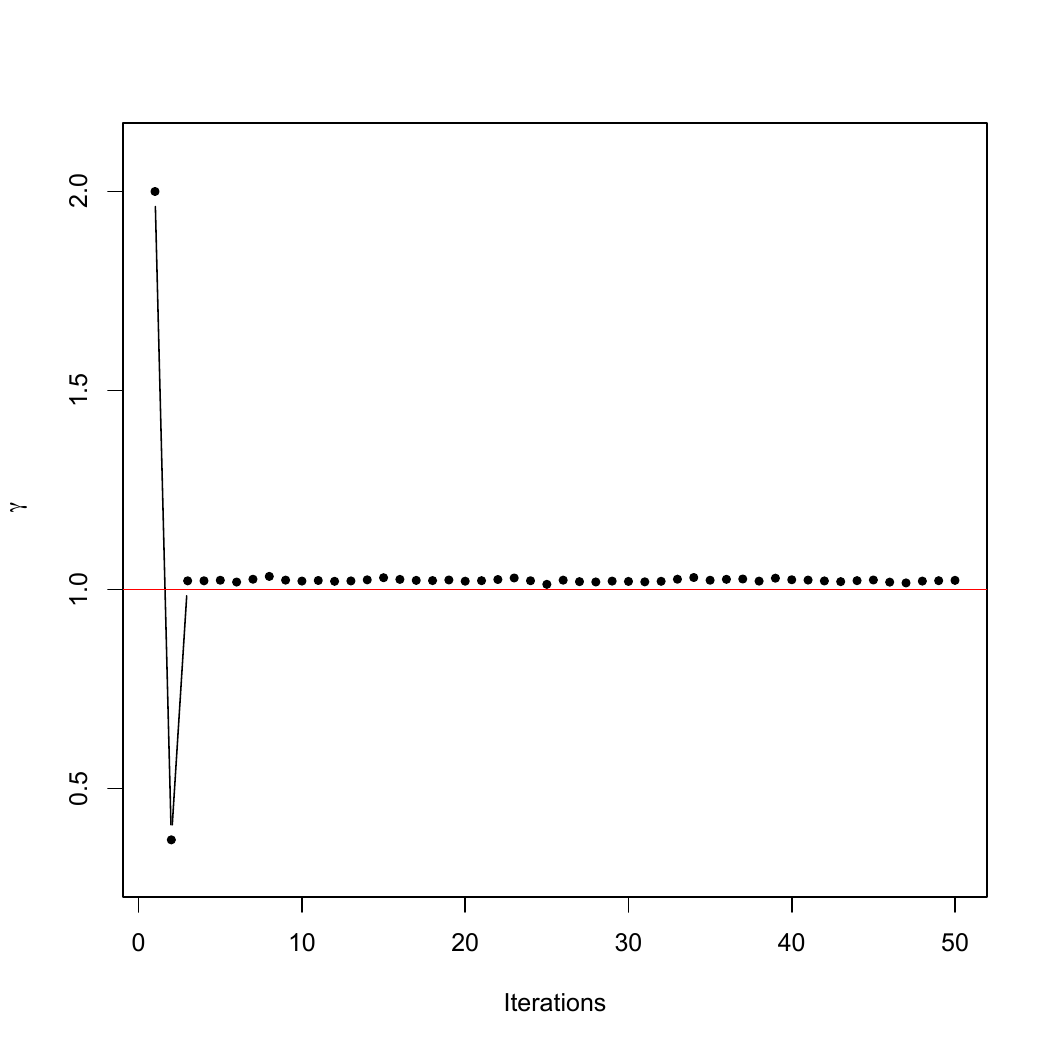}
\caption{\label{simulation_t2} 
The EM algorithm for the t-diffusion with random drift
parameter: parameter values in 50 iterations. The true parameter
values are $\beta = 0.1$ and  $\gamma =1$.} 
\end{center}
\end{figure}

We simulated 100 diffusions with $\beta = 0.1$ and $\gamma = 1$ at the
time points $t^i_j = j, j=1, \ldots, 100$. We ran 50 iterations of
the EM algorithm with initial values $\hat{\beta}_0 = 0.2$ and
$\hat{\gamma}_0 = 2$.
The parameter values in the iterations are plotted in Figure
\ref{simulation_t2}. The convergence was fast. As
estimators we use the final parameter values. The EM maximum likelihood
estimators  estimates are $\hat{\beta} = 0.1011$ and
$\hat{\gamma}=1.0098$.  

\sect{Application to neuronal data}
\label{neurodata}

In this section we estimate the parameters of a leaky
integrate-and-fire neuronal model (Ornstein-Uhlenbeck process) with
random level on the basis of measurement of the membrane potential
between firings. 

The data consists of measurements every 0.15 ms of the membrane
potential of a single auditory neuron of a guinea pig (for details on
the experiment, see \cite{exp}). When the potential crosses a certain
threshold, the neuron fires (produces an electric signal), resets to
an initial resting value and starts increasing towards a certain level
around which it fluctuates in a stationary way until the neuron fires
again. The data recorded
between firings (in so-called inter-spike intervals) can be considered
realisations of independent random processes with identical parameters
except that the stationary level varies randomly from one interval to
another. There are 312 inter-spike intervals, for which the number of
observations per interval, $n_i$, varies from hundreds to several
thousands.  The data was previously analysed by \cite{lansky}, who fitted
a fixed effect Ornstein-Uhlenbeck process individually to each inter-spike
interval. \cite{umberto1}, \cite{umberto2} and \cite{wiqvist} fitted an
Ornstein-Uhlenbeck process with a random level to the data using
methods different from ours, see the introduction, while \cite{dion2}
obtained a non-parametric estimate of the density of the random level.

The model is the Ornstein-Uhlenbeck process 
\[
dX^i_t = (a^i - \alpha X^i_t)dt + \beta dW^i_t,
\]
with random level $a^i$, where $\alpha > 0$, $\beta > 0$ and $a^i \sim
N(\xi, \sigma^2)$. We 
apply the Gibbs sampler, and as the prior distribution we take
$\alpha, \eta = \beta^{-2}, \xi$ and $\gamma = \sigma^{-2}$ to be independent and
exponential distributed with mean $\lambda_i^{-1}$, $i=1,2,3,4$,
respectively.

Since $f(x)=-x, g(x) = 1, h_\beta(x) = x/\beta$ and $\phi_{\alpha,
  \beta, a} (x) = -\alpha -2\alpha a x /\beta + \alpha^2 x^2 +
a^2/\beta^2$,  the quantities needed in the algorithm are 
\bean
t^i_{\alpha, \beta} &=& \beta^{-2} \left( x_{n_i}^i - x_1^i \right) +
\frac{\alpha}{\beta} \sum_{j=2}^{n_i} \int_{t^i_{j-1}}^{t^i_j} \left(  Y^{*ij}_s +
  \beta^{-1}\ell_1^{ij}(s)\right)^2 ds \\
B^i_\beta &=& \beta^{-2}  (t^i_{n_i} - t^i_1) \\
v_{\beta,\underline{\vect{a}}} &=& \sum_{i=1}^N \left[ \frac12 \beta^{-2} \left(
    (x_1^i)^2 - (x_{n_i}^i)^2 \right) +
\frac12 (t^i_{n_i} - t^i_1) + \beta^{-1} a^i
\sum_{j=2}^{n_i} \int_{t^i_{j-1}}^{t^i_j} \left(  Y^{*ij}_s +
  \beta^{-1}\ell_1^{ij}(s)\right) ds \right] \\
D_\beta &=& \sum_{i=1}^N \sum_{j=2}^{n_i} \int_{t^i_{j-1}}^{t^i_j} \left(  Y^{*ij}_s +
  \beta^{-1}\ell_1^{ij}(s)\right)^2 ds \\
G_1 &=& \frac12 \sum_{i=1}^N \sum_{j=2}^{n_i} \frac{(x_j^i -
  x_{j-1}^i)^2}{(t_j^i-t_{j-1}^i)} \\
G_2 &=& \sum_{i=1}^N \left[ \frac{\alpha}{2} \left( (x_{n_i}^i)^2 -
    (x_1^i)^2 \right) + a^i \left( x_{n_i}^i - x_1^i \right) 
\right] \\
E_1 &=& \frac12 \sum_{i=1}^N \left[ (a^i)^2  (t^i_{n_i} - t^i_1) +
  \alpha^2 \sum_{j=2}^{n_i} 
\int_{t^i_{j-1}}^{t_j^i} \ell^{ij}_1(s)^2 ds \right] \\
E_2 &=& \sum_{i=1}^N \left[ \alpha a^i \sum_{j=2}^{n_i}
  \int_{t^i_{j-1}}^{t^i_j} Y^{*ij}_s ds - \alpha^2 \sum_{j=2}^{n_i}
  \int_{t^i_{j-1}}^{t_j^i} Y_s^{*ij} \ell^{ij}_1(s) ds \right] \\
E_3 &=& - \alpha \sum_{i=1}^N a^i \sum_{j=2}^{n_i}
\int_{t^i_{j-1}}^{t_j^i} \ell^{ij}_1(s) ds 
\eean
where $\ell^{ij}_1(s)$ is given by (\ref{ell}). For this model
$F(\beta) = -\beta^{-2} (E_1 + E_3) + \beta^{-1} E_2$ (apart
from an additive term independent of $\beta$), but here we can make $E_1$
a part of the scale parameter of the weighted gamma distribution in
step 4 of the Gibbs sampler below, because $E_1$ is always positive.
The Gibbs sampler goes as follows.

\begin{description}

\item{0.} First draw $\alpha, \beta, \xi$ and $\gamma$ independently from the 
prior distribution, and draw $a_i$ from the normal distribution with
mean $\xi$ and variance $\gamma^{-1}$, independently for
$i=1,\ldots,N$.  
  
\item{1.} Simulate independent sample paths $Y^{*ij}$ conditionally on
  $a^i, \alpha, \beta$ and $X^i_{\mbox{\tiny obs}}$ for $j=2, \ldots,
  n_i, \ i=1, \ldots, N$ 

\item{2.} Draw $a_i$ with distribution $N((t^i_{\alpha, \beta} + \xi
  \gamma)/(B^i_\beta  + \gamma), 
  (B^i_\beta  + \gamma)^{-1})$, independently for $i=1,\ldots,N$

\item{3.}  Draw $\alpha$ from the distribution
  $N_+((v_{\beta,\underline{\vect{a}}}-\lambda_1)/D_\beta , D_\beta^{-1})$ 

\item{4.} Draw $\eta$ from the distribution with density function proportional to
  \[
\eta^{(n_\cdot-N)/2}\exp \left( - \eta (\lambda_2 + G_1 + E_1
  + G_2 + E_3) + \sqrt{\eta} E_2 \right), 
    \]
and set $\beta := \eta^{-1/2}$

\item{5.}  Draw $\xi$ from the $N(\bar a - \lambda_3/(\gamma N),
  (\gamma N)^{-1})$-distribution, where $\bar a = (a^1 +
  \cdots + a^N)/N$

\item{6.} Draw $\gamma$ from the $\Gamma ( N/2 +1, \lambda_4 + \frac12
  \sum_{i=1}^N (a^i - \xi)^2)$-distribution.

\item{7.} GO TO 1

\end{description}
If $\lambda_2 + G_1 + E_1 + G_2 + E_3>0$, the distribution in step 4 is a
weighted gamma distribution with scale parameter $(\lambda_2 + G_1 + E_1
+ G_2 + E_3)^{-1}$ and weight function $\exp (\sqrt{\eta} E_2)$.
If $\lambda_2 + G_1 + E_1 + G_2 + E_3 \leq 0$, the scale parameter is
$(\lambda_2 + G_1 + E_1)^{-1}$ (which is always positive) and the weight
function is $\exp (\sqrt{\eta} E_2 - \eta (G_2 + E_3))$. In step 1 we
simulated the approximate diffusion bridges of \cite{bladtsorensen},
and in step 4 we used approximate direct sampling, see Subsection
\ref{expGibbs}.

\begin{figure}
\begin{center}
\includegraphics[width=4.0cm]{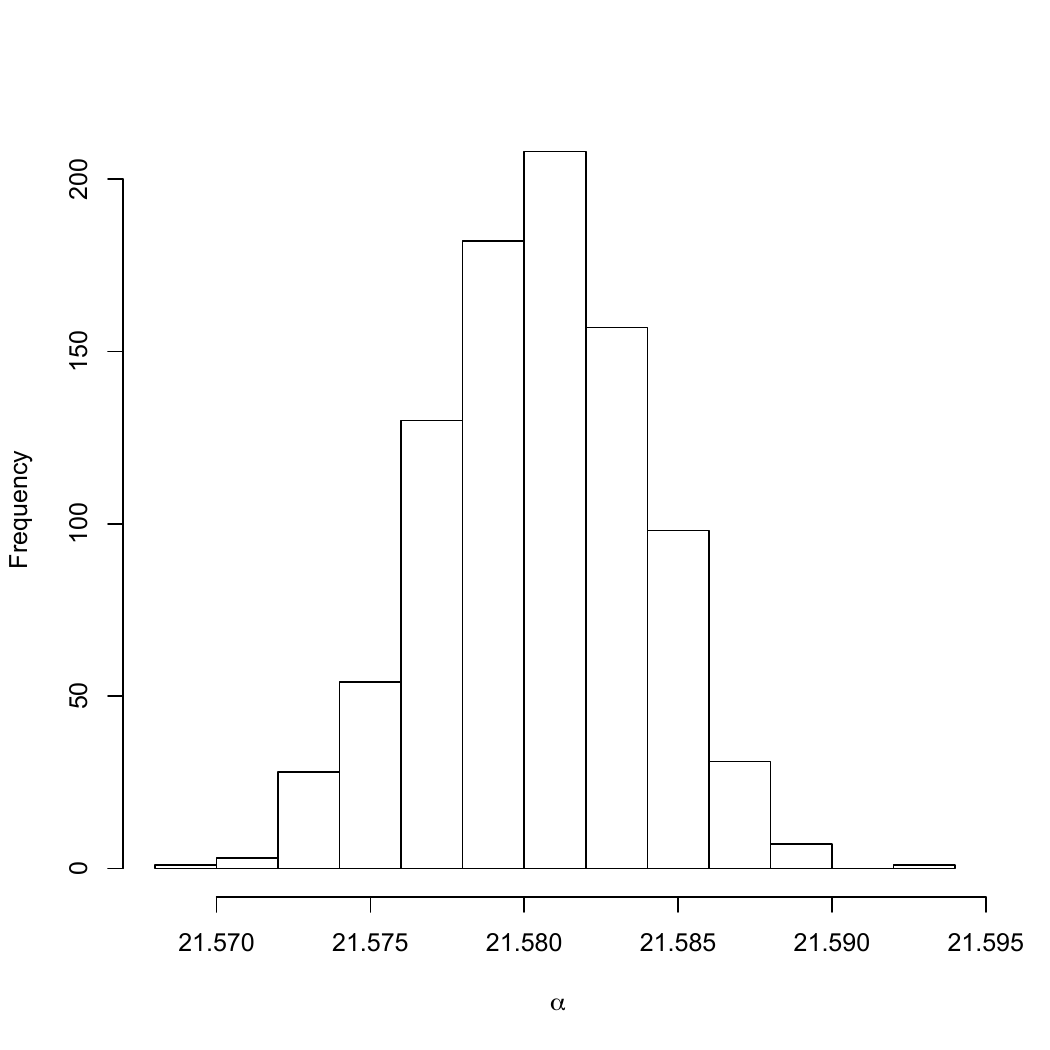} 
\includegraphics[width=4.0cm]{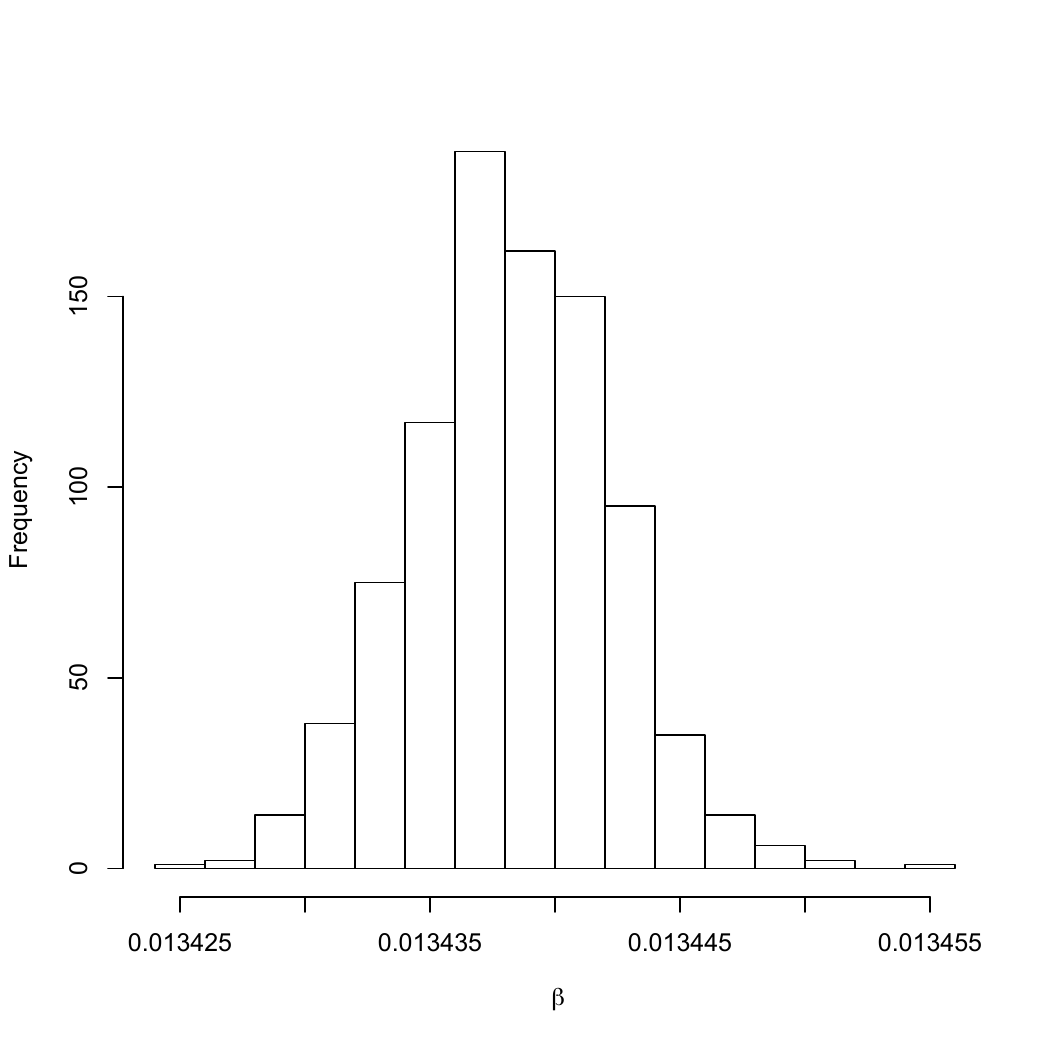}
\includegraphics[width=4.0cm]{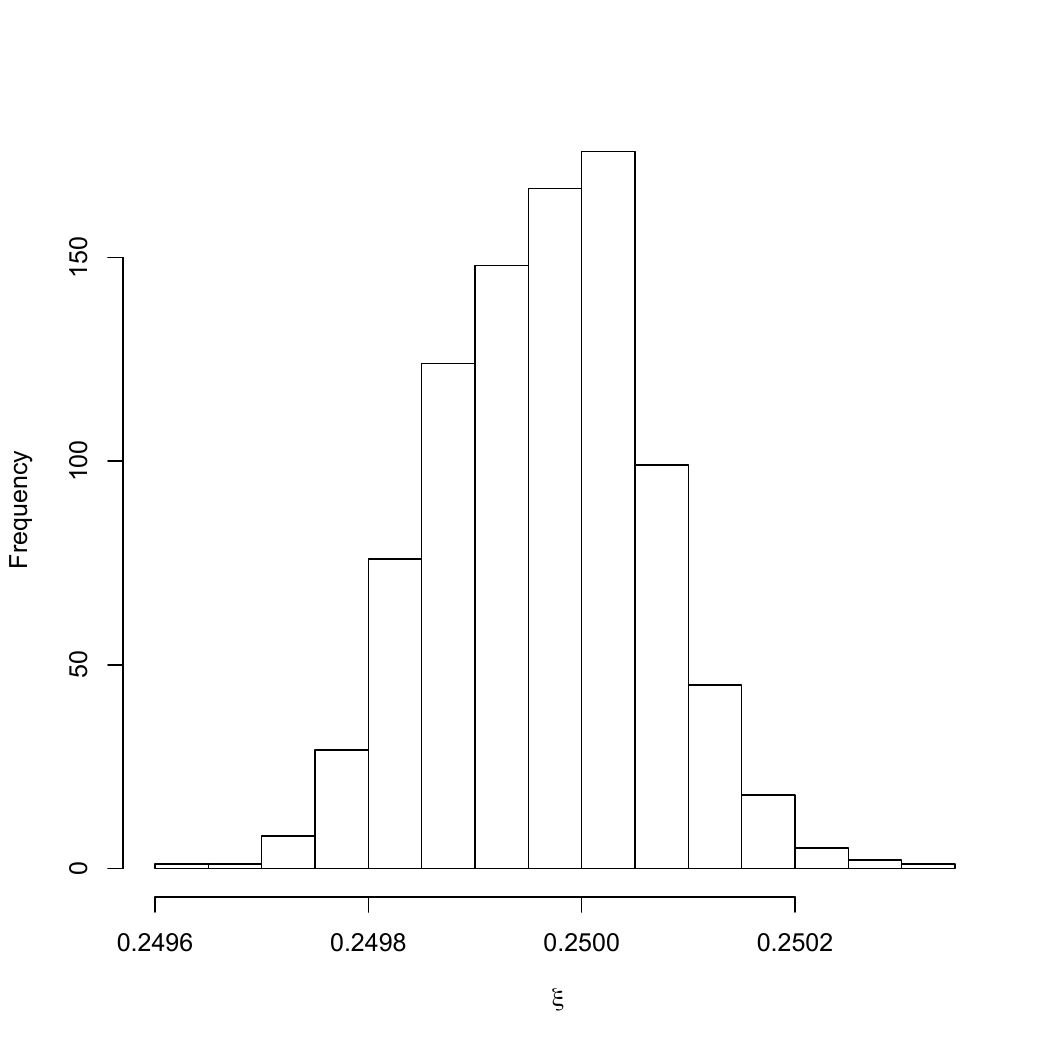}
\includegraphics[width=4.0cm]{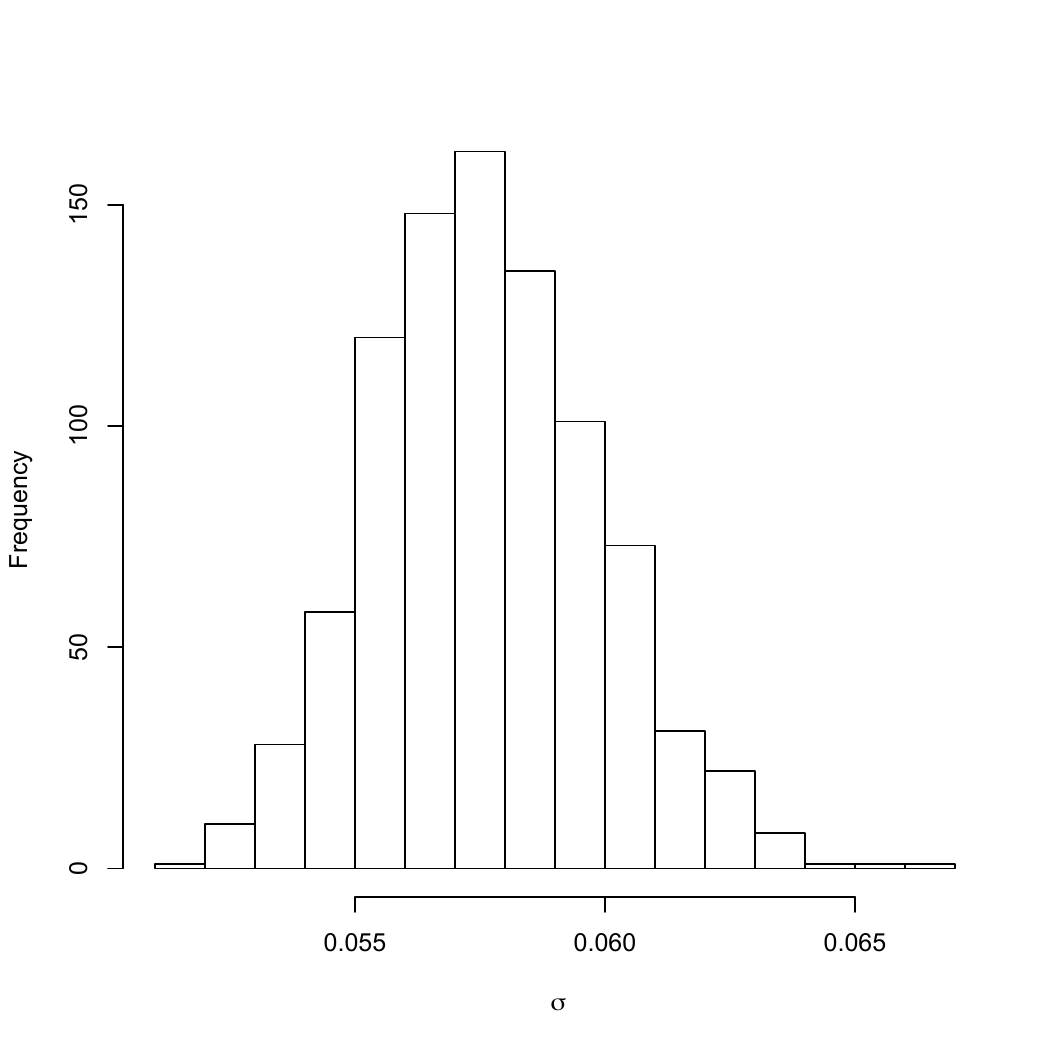}
\caption{\label{neurofig} 
Histograms of the Gibbs sampler draws of the parameters $\alpha$,
$\beta$, $\xi$ and $\gamma$ of the neuron model (after a
burn-in of 100 draws).} 
\end{center}
\end{figure}

The Gibbs sampler was run with 1000 iterations using prior
distributions with parameters $\lambda_1 = 5, \lambda_2 = 0.02,
\lambda_3 = 400, \lambda_4 = 0.35$, which are expected to be
essentially non-informative. After a burn-in of 100 iterations,
the estimates are based on the last 900 draws of the parameters. 
Histograms of the last 900 parameter draws are plotted in
Figure \ref{neurofig}. The
results, see Table \ref{neurotable}, are in reasonable accordance with the
estimates obtained in \cite{lansky}, \cite{umberto2},
\cite{wiqvist} and \cite{dion2}. The estimates of $\beta$ are
essentially the same, while the estimate of $\sigma$ is close to the
mean of the estimates in the other papers (which vary quite a bit
between the papers). \cite{umberto2} are unhappy with their own
estimate of $\alpha$ and prefer a value from the literature, namely
25.6, which is close to our estimate.    

\begin{table}
\begin{center}
\begin{tabular}{|c|c|c|c|c|}
\hline
Parameter & mean & 0.025-quantile & 0.975-quantile \\
\hline
$\alpha$ & 21.580 & 21.574  & 21.586 \\
\hline
 $\beta$ & 0.013438 & 0.013431 & 0.013446 \\
\hline
 $\xi$ & 0.2500 & 0.2498 & 0.2502 \\
\hline 
$\sigma$ & 0.0576 & 0.0536 & 0.0624 \\
\hline 
\end{tabular}

\end{center}
\caption{\label{neurotable}
  Estimates obtained by 900 iterations of the Gibbs sampler
  after a burn-in of 100 iterations.} 
\end{table}

\appendix

\renewcommand{\theequation}{\Alph{section}.\arabic{equation}}

\sect{Appendix: Simple diffusion bridge simulation}
\label{appen}

In this appendix we briefly present the simple method of simulating
diffusion bridges introduced in \cite{bladtsorensen} and the
corrigendum \cite{bladtmidersorensen}, which is key to the algorithms
proposed in this paper. 

The aim is to simulate a $(t_1, x_1, t_2, x_2)$-bridge for the
diffusion process given by (\ref{basicmodel}), which is assumed to be
ergodic with invariant probability density function $\nu$
w.r.t. Lebesgue measure on the state space. The method is based on the
following simple construction of a process that starts from $x_1$ at
time zero and at time $t_1$ ends at $x_2$, which is a good
{\it approximation to a $(t_1, x_1, t_2, x_2)$-bridge}. One diffusion process,
$X_t$, that solves (\ref{basicmodel}), is started from the point 
$x_1$, while another independent solution to  (\ref{basicmodel}),
$\bar{X}_t$, is started from the point $x_2$. The time of the second
diffusion is then reversed to obtain the process $X'_t = \bar{X}_{t_2-t}$
Suppose there is a time point $\tau \in [t_1,t_2]$ at which
$X_\tau = X'_\tau$. Then the process that is equal to
$X_t$ for $t \in [t_1,\tau]$ and for $t \in [\tau,t_2]$ equals $X'_t$
is a process that starts at $x_1$ and ends at $x_2$. Because the probability that $X$
and $X'$ meet in $[t_1,t_2]$ is considerable, provided $t_2-t_1$ is not very
small, such a process can be obtained by rejection sampling and can be
shown to be an approximation to a $(t_1, x_1, t_2, x_2)$-bridge in  a
sense explained in \cite{bladtmidersorensen}. 

In practice, the approximate diffusion bridge is simulated as
follows. Let $Y_{\delta i}$, $i=0,1, \ldots, N$ and $\bar{Y}_{\delta
  i}$, $i=0,1, \ldots, N$ be (independent) simulations, for instance
using the Eule scheme, of $X$ and $\bar{X}$ in $[t_1,t_2]$ 
with step size $\delta = (t_2-t_1)/N$. Then a simulation of an
approximation to a $(t_1,x_1,t_2,x_2)$-bridge is obtained by the
following rejection sampling algorithm. Keep simulating $Y$ and
$\bar{Y}$ until there is an $i$ such that either $Y_{\delta i} \geq
\bar{Y}_{\delta (N-i)}$ and $Y_{\delta (i+1)} \leq
\bar{Y}_{\delta(N-(i+1))}$ or $Y_{\delta i} \leq \bar{Y}_{\delta (N-i)}$ and 
$Y_{\delta (i+1)} \geq \bar{Y}_{\delta (N-(i+1))}$. Once this has been achieved, define
\[
B_{\delta i}=\left\{ 
\begin{array}{ll}
Y_{\delta i} & \mbox{\rm for } \ i = 0,1, \ldots, \mu-1 \\ & \\
\bar{Y}_{\delta (N-i)} & \mbox{\rm for } \ i=\mu, \ldots N, 
\end{array}
\right.
\]
where $\mu = \min \{ i \in \{1, \ldots N\} | Y_{\delta i} \leq
\bar{Y}_{\delta (N-i)} \}$ if $Y_0 \geq \bar{Y}_\Delta$, and 
$\mu = \min \{ i \in \{1, \ldots N\} | Y_{\delta i} \geq
\bar{Y}_{\delta (N-i)} \}$ if $Y_0 \leq \bar{Y}_\Delta$. Then $B$ approximates
a $(t_1,x_1,t_2,x_2)$-bridge. Apart from the usual discretization
error, the step size $\delta$ also controls the probability that a
trajectory crossing is not detected. Therefore, it is advisable to
choose $\delta$ smaller than usual.

In several cases the approximate diffusion bridge is a sufficiently
good approximation to be used as it is, as we do in the simulation
studies and data applications in this paper. However, an {\it exact
diffusion bridge} (appart from the discretization error) can be
obtained via a pseudo-marginal Metropolis-Hastings algorithm with
target distribution equal to that of an exact diffusion bridge, see
\cite{andrieuroberts}. In the $k$th step of the algorithm, we simulate
as proposal an approximate diffusion bridge $X^{(k)}$ as explained
above supplemented by a geometric random variable $S^{(k)}$, which is obtained
as follows. Simulate a sequence of independent solutions $Z^{(i)}$ to
(\ref{basicmodel}) in $[t_1,t_2]$ with $Z_{t_1}^{(i)} \sim \nu$ until a sample path
is obtained that intersects $X^{(k)}$ in $[t_1,t_2]$. Then $$S^{(k)}
= \min \{ i \,  : \,  Z^{(i)} \mbox{ intersects } X^{(k)} \}. $$ 
The proposed diffusion bridge $X^{(k)}$ is accepted with probability
$\min \{ 1, S^{(k)}/S^{(k-1)}   \}$. We call $S^{(k)}$ the {\it geometric
variable associated with} $X^{(k)}$. In order to reduce the variance,
we could alternatively replace $S^{(k)}$ by the average $T^{(k)}$ of a
number of independent geometric random variables associated with
$X^{(k)}$, but if the probability that $Z^{(i)}$ intersects $X^{(k)}$
is small, this might be time consuming. 

Suppose the distribution of the diffusion process depends on a
parameter vector $\vect{\theta}$ with prior distribution $\pi
(\vect{\theta})$, and that we want to use the simple bridge simulation
method in a Gibbs sampler that alternates between drawing $\vect{\theta}$
conditional on a $(t_1, x_1, t_2, x_2)$-bridge $X$ and drawing $X$
conditional on $\vect{\theta}$. If we use the approximate bridge, this is
simple to do. If we want to simulate exact bridges, we can use a
{\it Metropolis within Gibbs} algorithm. In the $i$th iteration we draw
$\vect{\theta}_i$  and $X_i$. Conditional on $\vect{\theta}_i$, we draw
an approximate bridge $X_{a,i}$ and its associated geometric variable
$S_i$. With probability $\min \{ 1, S_i/S_{i-1} \}$ we accept the
proposed value, i.e.\ $X_i := X_{a,i}$. Otherwise,  $X_i := X_{i-1}$.

Everything described above can be generalized to multivariate
diffusion processes, see \cite{bladtfinchsorensen,corbladtfinchsorensen}.
The construction of the approximate diffusion bridge and simulation of
the associated geometric variables are more complicated in the multivariate case
because coupling methods for diffusions must be applied to ensure that
sample paths intersect with positive probability. The methods for
multivariate diffusions can also be used to improve the computational
efficiency of the simulation of one-dimensional diffusion bridges. 

\section*{Acknowledgement}

The work of F. Baltazar-Larios was supported by UNAM-DGAPA-PAPIIT-IN102224.

\bibliographystyle{natbib}
\bibliography{randomeffect3}

\end{document}